\documentclass[aps,prl,twocolumn,superscriptaddress,floatfix,showpacs,groupedaddress]{revtex4-1}
\usepackage{hyperref}
\usepackage{graphicx}
\usepackage{amsfonts}
\usepackage{amssymb}
\usepackage{amsmath}
\usepackage{txfonts}
\usepackage{lipsum}
\usepackage{color}
\usepackage{wasysym}

\usepackage{letltxmacro}
\LetLtxMacro{\oldsqrt}{\sqrt}
\renewcommand{\sqrt}[2][\mkern8mu]{\mkern-6mu\mathop{}\oldsqrt[#1]{#2}}

\newcommand{\av}[1]{\ensuremath{\left\langle #1 \right\rangle}}
\newcommand{\qv}{\mathbf{q}}
\newcommand{\kv}{\mathbf{k}}

\begin{document}
\title{
Dynamical and reversible control of topological spin textures\\
}
\author{E. A. Stepanov$^1$, C. Dutreix$^{1,2}$, M. I. Katsnelson$^1$}
\affiliation{$^1$Radboud University, Institute for Molecules and Materials, Heyendaalseweg 135, 6525AJ Nijmegen, Netherlands\\
$^2$Univ Lyon, Ens de Lyon, Univ Claude Bernard, CNRS, Laboratoire de Physique, F-69342 Lyon, France}

\begin{abstract}
Recent observations of topological spin textures brought spintronics one step closer to new magnetic memories. Nevertheless, the existence of Skyrmions, as well as their stabilization, require very specific intrinsic magnetic properties which are usually fixed in magnets. Here we address the possibility to dynamically control their intrinsic magnetic interactions by varying the strength of a high-frequency laser field. It is shown that drastic changes can be induced in the antiferromagnetic exchange interactions and the latter can even be reversed to become ferromagnetic, provided the direct exchange is already non-negligible in equilibrium as predicted, for example, in Si doped with C, Sn, or Pb adatoms. In the presence of Dzyaloshinskii-Moriya interactions, this enables us to tune features of ferromagnetic Skyrmions such as their radius, making them easier to stabilize. Alternatively, such topological spin textures can occur in frustrated triangular lattices. Then, we demonstrate that a high-frequency laser field can induce dynamical frustration in antiferromagnets, where the degree of frustration can subsequently be tuned suitably to drive the material toward a Skyrmionic phase.
\end{abstract}

\maketitle

In the 1960s, Skyrme solved the equation of motion for a linear sigma model Lagrangian and reported static classical solutions which are now referred to as Skyrmions~\cite{SKYRME1962556}.
Remarkably, the boundary conditions they satisfy allow them to be characterized by a topological charge. The elementary particles they described were identified as three-quark-made objects, namely, baryons, the family to which belong protons and neutrons.
Skyrmions were later predicted in condensed matter physics too, as nontrivial spin textures~\cite{bogdanov1989thermodynamically}. Importantly, this prediction has recently been confirmed experimentally by neutron scattering in three-dimensional helical magnets MnSi~\cite{Muhlbauer915} and Fe$_{1-x}$Co$_{x}$Si~\cite{PhysRevB.81.041203}, by electron microscopy in two-dimensional helical magnet Fe$_{0.5}$Co$_{0.5}$Si~\cite{yu2010real}, and by spin-polarized scanning tunneling microscopy in Fe films deposited onto the Ir(111) surface ~\cite{heinze2011spontaneous}. The observations of such topological magnetic structures have been a decisive step forward in the perspective of Skyrmion-based data storage in spintronics~\cite{Jonietz1648,fert2013skyrmions,Romming636,Tomasello:2014qd}.
From a fundamental viewpoint, skyrmions arise from different mechanisms. They appear in thin films under perpendicular magnetic field due to the competition between an easy-axis anisotropy and dipolar interactions that, respectively, favor out-of- and in-plane magnetizations~\cite{lin1973bubble,malozemoff1979magnetic,garel1982phase}. If they were also observed as a result of four-spin exchange interactions~\cite{heinze2011spontaneous}, this is within the context of frustrated exchange interactions (FEI) ~\cite{PhysRevLett.108.017206,leonov2015multiply,PhysRevB.93.064430,PhysRevB.93.184413} and Dzyaloshinskii-Moriya interactions (DMI)~\cite{DZYALOSHINSKY1958241,PhysRev.120.91,pfleiderer2011magnetic,shibata2013towards,emori2013current,ryu2013chiral,ryu2014chiral,franken2014tunable,PhysRevB.88.214401,PhysRevB.90.020402,chen2013tailoring} that Skyrmions are mainly discussed nowadays. In noncentrosymmetric ferromagnets DMI compete with the exchange interactions to yield a helical spiral phase which, under an external magnetic field, may lead to a Skyrmionic phase~\cite{PhysRevB.82.094429, PhysRevX.4.031045, PhysRevB.92.214439, PhysRevB.92.134405, 2016arXiv160303688K, 1367-2630-18-6-065003}. Nevertheless, DMI-based Skyrmions have a broad size, basically 5--100 nm, which makes them hard to stabilize~\cite{nagaosa2013topological,PhysRevX.4.031045}.
In frustrated magnets, FEI-based Skyrmions may also arise from the competition between ferromagnetic (FM) nearest-neighbor (NN) and antiferromagnetic (anti-FM) next-NN exchange interactions~\cite{PhysRevLett.108.017206, leonov2015multiply, PhysRevB.93.064430,PhysRevB.93.184413}. However, the Skyrmionic phase additionally requires very special strengths for these two interactions. Thus, the main difficulty with controlling FEI- and DMI-based Skyrmions relies on the intrinsically fixed magnetic properties of materials. Tuning and controlling FEI and DMI then becomes extremely challenging.  Research in this direction has recently been undertaken, thus reporting the possibility to tune DMI via anisotropy~\cite{emori2013current, ryu2013chiral, ryu2014chiral, franken2014tunable, PhysRevB.88.214401, PhysRevB.90.020402, chen2013tailoring}, hydrostatic pressure~\cite{0953-8984-17-10-018, ritz2013formation, PhysRevB.87.134424}, or mechanical strain~\cite{shibata2015large}.

Here, we report the possibility to dynamically control the intrinsic magnetic interactions by varying the strength of a high-frequency laser field, and subsequently tune the Skyrmionic features they are responsible for. The idea simply relies on the fact that DMI and FEI are both based on hopping processes, and that time-periodic fields renormalize the electronic tunneling, leading to phenomena such as dynamical Wannier-Stark localization \cite{dunlap1988effect}, symmetry-protected topological transitions~\cite{Oka:2009mz, kitagawa2010topological, lindner2011floquet, Carpentier:2015qy, PhysRevB.93.241404, cayssol2013floquet}, or ultrafast control of magnetism \cite{PhysRevB.90.214413, NatCom1, NatCom2, PhysRevLett.117.147202}. Here, we show that drastic changes can be induced in the antiFM exchange interactions that can even be switched to FM, provided the direct exchange interaction is already reasonable in equilibrium. This dynamical anti-FM -- FM phase transition is predicted in Si(111) doped with Sn or Pb adatoms under infrared light. Moreover, DMI are also renormalized by the laser field, which allows to dynamically tune features of FM Skyrmions such as their radius, making them easier to stabilize. In the case of FEI in triangular lattices, anti-FM Skyrmions have been predicted too, but no suitable magnets are available for experimental realizations so far. Then we suggest a possible route to induce dynamical frustration in antiferromagnets, and subsequently drive the degree of frustration until the material enters a Skyrmionic phase. Possible applications of this prescription are finally discussed in materials such as C$_2$F and Si(111) doped with C adatoms.\\

{\it{Skyrmion model with DMI}} --
Let us start with the following tight-binding Hamiltonian
\begin{align}
H &= \sum_{\av{ij},\,\sigma\sigma'}c^{*}_{i\sigma}\left(t\,\delta^{\phantom{*}}_{\sigma\sigma'} + i\boldsymbol{\Delta}^{\phantom{*}}_{ij}\,
\boldsymbol{\sigma}^{\phantom{*}}_{\sigma\sigma'}
\right)c^{\phantom{*}}_{j\sigma'} 
+ \sum_{i}U^{\phantom{*}}_{00}\,
n^{\phantom{*}}_{i\uparrow}n^{\phantom{*}}_{i\downarrow}
\label{eq:Hamiltonian1}\\
&+\frac12\sum_{\av{ij},\,\sigma\sigma'} 
U^{\phantom{*}}_{\av{ij}}\,
n^{\phantom{*}}_{i\sigma}n^{\phantom{*}}_{j\sigma'}
-\frac12\sum_{\av{ij},\,\sigma\sigma'}
J^{\rm D}_{\av{ij}}\,
c^{*}_{i\sigma}c^{\phantom{*}}_{i,\sigma'}
c^{*}_{j,\sigma'}c^{\phantom{*}}_{j\sigma}, \notag
\end{align}
where $t$ denotes the NN hopping amplitudes of electrons on a triangular lattice. Vector $\boldsymbol{\Delta}^{\phantom{*}}_{ij}=\left(\Delta^{x\phantom{y}}_{ij}, \Delta^{y}_{ij}, 0\right)=-\boldsymbol{\Delta}^{\phantom{*}}_{ji}$ describes the Rashba spin orbit and lies perpendicularly to the bond between NN sites $i$ and $j$, while $\boldsymbol{\sigma}^{\phantom{*}}_{\sigma\sigma'}=\left(\sigma^{x}, \sigma^{y}, \sigma^{z}\right)$ is a vector of Pauli matrices. Besides, $U_{00}$ and $U_{\av{ij}}$ refer to onsite and NN Coulomb interactions, and $J^{D}_{\av{ij}}$ is the NN FM direct exchange interaction. The latter can be comparable to the anti-FM kinetic exchange interaction in LiCu$_{2}$O$_{2}$, SrCu$_{2}$(BO$_{2}$)$_{2}$ and Si(111) with adatoms~\cite{PhysRevB.75.224408, PhysRevB.78.195110, 2016arXiv160907648B}, and even compensate it in C$_2$F~\cite{Rudenko, PhysRevB.88.081405}. The direct DMI is usually small and may even vanish in some 2D materials due to symmetry arguments~\cite{PhysRevB.78.195110}; thus, it is disregarded here.\\

{\it{High-frequency description}} --
Now we aim to provide an effective description of the system when electrons are rapidly driven by a time-periodic laser of frequency $\Omega$. The vector potential it leads to in the temporal gauge is ${\bf A} =~\left(A_{x}\,a_{0}\cos\left(\Omega{}t\right), A_{y}\,a_{0}\sin\left(\Omega{}t-\phi\right), 0\right)$, where $a_{0}$ is the lattice constant and $c=\hbar=1$. It is described via Peierls substitution $\kv\to\kv-e{\bf A}(t)$ in the momentum representation of the Hamiltonian, where $e$ denotes the electron charge. Phase $\phi$ characterizes the light polarization that is elliptic for $\phi=0$ and linear for $\phi=\pi/2$. The Hamiltonian becomes time periodic and its quantum nonequilibrium steady states obey the time-dependent Schr\"odinger equation $i \partial_\tau \Psi(\lambda, \tau) =~\frac{1}{\Omega} H(\tau) \Psi (\lambda, \tau)$, where $\tau=\Omega t$. Here, we have introduced a dimensionless parameter $\lambda=\delta E/ \Omega$ that compares a certain energy scale $\delta E$ to the field frequency. For simplicity, we chose $\delta E$ as the largest energy scale involved in Hamiltonian \eqref{eq:Hamiltonian1} among $t_{ij}$, $|\boldsymbol{\Delta}^{\phantom{*}}_{ij}|$, $U_{ij}$, and $J^{\rm D}_{ij}$.
Then, the Schr\"odinger equation reads
$i \partial_\tau \Psi(\lambda, \tau) =~\lambda\overline{H}(\tau) \Psi (\lambda, \tau)$, where the Hamiltonian is now renormalized as $\overline{H}(\tau)=H(\tau)/\delta{}E$. In the high-frequency limit, $\lambda$ is small and we can look for a unitary transformation defined as $\Psi (\lambda, \tau) =~\exp\{-i\Delta(\tau)\}\,\psi (\lambda, \tau)$, which removes the time dependence of the Hamiltonian~\cite{Itin1,Itin2}. By construction we also impose $\Delta(\tau) =~ \sum_{n=1}^{+\infty} \lambda^{n} \Delta_{n}(\tau)$, with $\Delta_{n}(\tau)$ a $2\pi$ periodic function that averages at zero. Such a transformation leads to $i\partial_\tau \psi(\lambda, \tau) = \frac{1}{\Omega}{\cal H} \psi (\lambda, \tau) = \lambda \overline{\cal H} \psi (\lambda, \tau)$, where
$\overline{\cal H}=~\sum_{n=0}^{+\infty}\lambda^{n}\tilde{H}_{n}$. Then $\tilde H_{n}$ and $\Delta_{n}$ are determined iteratively in all orders in $\lambda$ (see, e.g., Refs.~\cite{PhysRevB.93.241404,SM}). Here, we restrict the analysis to the second order in $\lambda$. The effective time-independent Hamiltonian it leads to is
\begin{align}
{\cal H} &= \sum_{\av{ij},\,\sigma\sigma'}c^{*}_{i\sigma}\left(t'
\,\delta^{\phantom{*}}_{\sigma\sigma'} + i
\boldsymbol{\Delta}'_{ij}\,
\boldsymbol{\sigma}^{\phantom{*}}_{\sigma\sigma'}
\right)c^{\phantom{*}}_{j\sigma'} 
+\sum_{i}U_{00}\,
n^{\phantom{*}}_{i\uparrow}n^{\phantom{*}}_{i\downarrow}\\
&+\frac12\sum_{\av{ij},\,\sigma\sigma'}
U_{\av{ij}}\,
n^{\phantom{*}}_{i\sigma}n^{\phantom{*}}_{j\sigma'}
-\frac12\sum_{ij,\,\sigma\sigma'}
J'^{\rm D}_{\av{ij}}\,
c^{*}_{i\sigma}c^{\phantom{*}}_{i,\sigma'}
c^{*}_{j,\sigma'}c^{\phantom{*}}_{j\sigma}. \notag
\end{align}
Kinetic hopping and the Rashba spin orbit are both NN hopping processes, so they are both renormalized in the same way by the laser field: $t' = t\,{\cal J}_0(Z)$ and $\boldsymbol{\Delta}'_{ij} =~\boldsymbol{\Delta}'_{ij}\,{\cal J}_0(Z)$, where ${\cal J}_0$ is the 0th order Bessel function, the polarization is assumed to be circular ($\phi=0$), and $Z = eA_{x}a_{0}=eA_{y}a_{0}=eE_{0}a_{0}/\Omega$ with $E_{0}$ the laser field strength. The explicit expression of the renormalized direct exchange interaction $J'^{D}_{\av{ij}}$ is provided in Ref.~\cite{SM}. Importantly, the effective Hamiltonian derived above from the high-frequency expansion remains a good approximation of the dynamics over a time scale ${\cal T}_{\rm heating}\sim{}{\rm exp}[{\cal O}(\lambda^{-1})]$ that is exponentially long with the frequency~\cite{Kuwahara201696,2016arXiv161105024H}, and during which heating can be neglected. Indeed, the time scale after which the heating of the system becomes crucial is much larger than the measurement time ${\cal T}_{\rm heating}\gg{}mT$. Here, $m$ is the number of driving periods $T= 2\pi/\Omega$~\footnote{The electron-electron interaction introduced in our model additionally plays the role of relaxation mechanism that allows to obtain such a nonequilibrium steady states~\cite{glazman1981resonant, glazman1983kinetics}. The detailed investigation of the processes that help electrons to relax into such a stabilized regime is currently under investigations~\cite{PhysRevLett.112.050601, PhysRevX.6.021022, PhysRevX.5.041050, PhysRevE.93.012130, PhysRevLett.116.120401}.}.

In the strong localization regime ($t_{ij}\ll{}U_{00}$), one can construct a Heisenberg Hamiltonian in terms of spin operators $\hat{\bf S}_{i}$ and superexchange as proposed by Anderson~\cite{PhysRev.115.2} and Moriya~\cite{PhysRev.120.91}
\begin{align}\label{Spin Hamiltonian}
&H_{\rm spin} = -\sum_{\av{ij}}J_{ij}\,\hat{\bf S}_{i}\,\hat{\bf S}_{j} +
\sum_{\av{ij}}{\bf D}_{ij}\,[\hat{\bf S}_{i}\times\hat{\bf S}_{j}].
\end{align}
Here, ${\bf D}_{ij} =4t'{\bf \Delta}_{ij}'/\tilde{U}$, where $\tilde{U}=~U_{00}-U_{\av{ij}}$~\cite{PhysRevLett.111.036601}. It characterizes DMI, namely, antisymmetric anisotropic interactions that are responsible for the weak ferromagnetism of some antiferromagnets \cite{DZYALOSHINSKY1958241,PhysRev.120.91,PhysRevB.52.10239}. This interaction scales with ${\cal J}_0^{2}(Z)$. Importantly, no additional contribution to DMI can be effectively induced by the high-frequency light \cite{SM} and, therefore, DMI cannot change signs when varying the field strength. Besides, there may be a third term in Eq.\,(\ref{Spin Hamiltonian}) which, as introduced in Moriya's seminal paper \cite{PhysRev.120.91}, describes symmetric anisotropic interactions. Nevertheless, it scales with $\Delta_{ij}'^{2}$ and since $\Delta_{ij}'^{2} \ll t'_{ij}\Delta'_{ij}$, this term can safely be neglected for all strengths of the laser field \cite{SM}. Note that, finally, a Zeeman magnetic field $h$ could also be included in Hamiltonian (\ref{Spin Hamiltonian}) through $\sum_{i}\hat{S}_{i}^{z}\,h$ as in Refs.~\cite{PhysRevB.92.214439, PhysRevB.93.184413}. However, it would neither be renormalized by the high-frequency field, nor be responsible for any correction up to the second order in the high-frequency expansion \cite{SM}. This is the reason why it is disregarded here. The isotropic symmetric exchange interaction between two spins satisfies
$J=J^{ D} + J^{ D}_{\rm ind}-J^{ K}$. Here, $J^{ D}$ denotes the direct exchange interactions which takes place in the material in equilibrium, i.e., in the absence
of the laser field. The anti-FM kinetic exchange interaction $J^{ K} = 2t^{2}{\cal J}_{0}^{2}(Z)/\tilde{U}$ already exists in equilibrium, but it is renormalized by the field strength. Finally $J^{ D}_{\rm ind}\simeq 4t^{2}\tilde{U}{\cal J}_{1}^{2}(Z)/\Omega^2$ is a FM field-induced correction to the direct exchange and is a purely nonequilibrium effect. 

\begin{figure}[!t]
\centering
\includegraphics[width=1\linewidth]{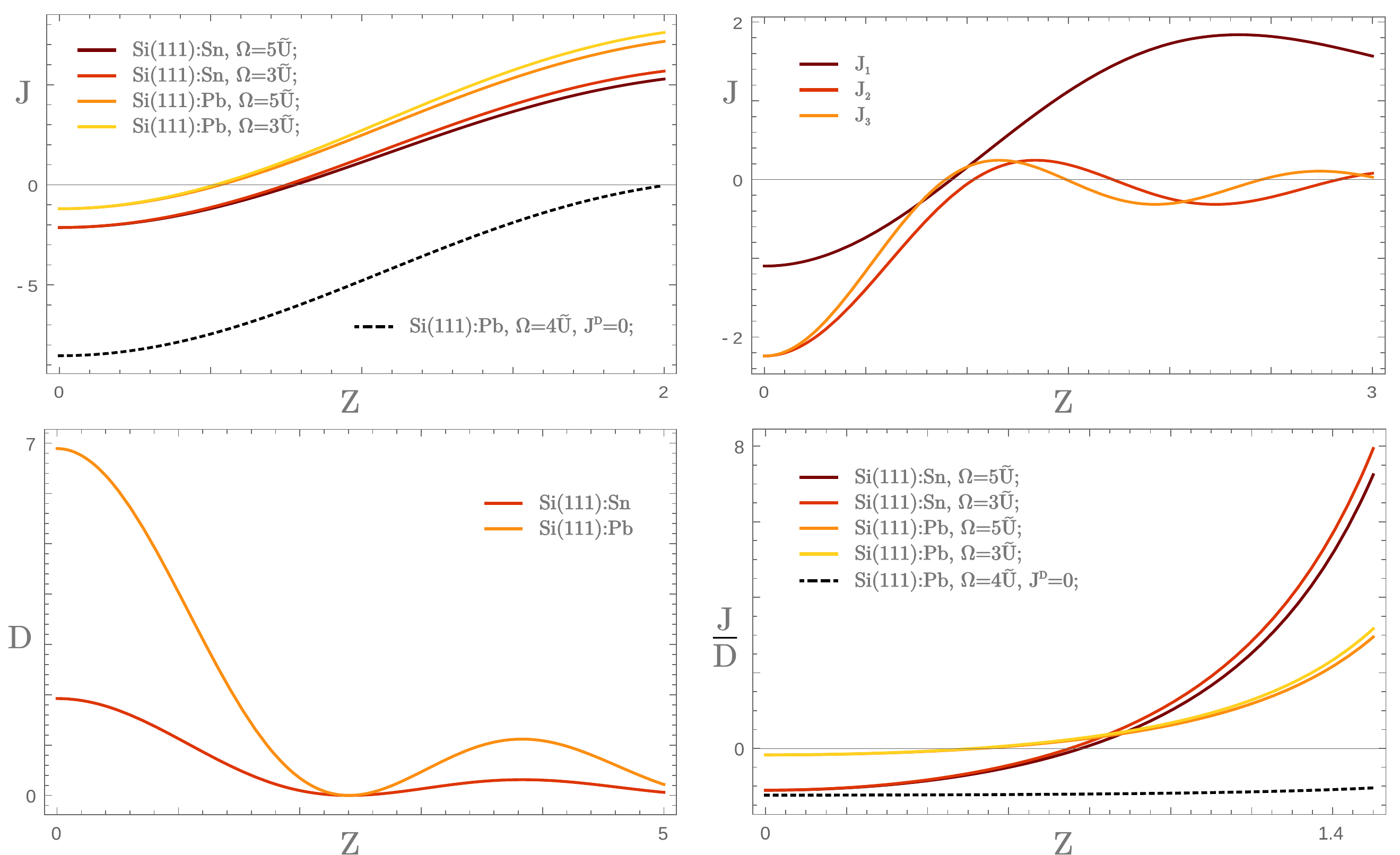}
\caption{(color online) Magnetic properties of the Si(111):\{Sn, Pb\} systems as functions of the laser field strength $Z$ for different frequencies $\Omega=3\tilde{U}, 5\tilde{U}$: exchange interactions $J$ (top left), DMI (bottom left), ratio $J/D$, which is proportional to a Skyrmion radius (bottom right). Dashed black curve corresponds to the case of zero direct exchange and shows the important role that $J^{ D}$ plays in a phase transition and manipulation of the Skyrmionic structure. Top right panel shows NN and next-NN exchange interactions of Si(111):C, where we take ``unrealistic'' case of $t_1=t_2=t_3$ to make the difference in anti-FM -- FM transition more visible. All units are given in meV.}
\label{fig:exchange-radius}
\end{figure}

If the Hubbard Hamiltonian that leads to Eq. (\ref{Spin Hamiltonian}) lies in an anti-FM phase in equilibrium, it is remarkable that it undergoes a dynamical phase transition to become FM when varying the field strength out of equilibrium. This is illustrated by the positive values of exchange interaction $J$ in Fig.\,\ref{fig:exchange-radius}. Because of field-induced correction $J^{ D}_{\rm ind}$, this dynamical transition is even predicted to occur when the direct exchange $J^{ D}$ is absent in equilibrium, as in iron oxides \cite{NatMaz}. When $J^{ D}=0$, the transition roughly requires $\tilde{U}\sim\Omega$ and, additionally, $eE_{0}a_{0}\sim 2\Omega$ according to Fig.\,\ref{fig:exchange-radius}. Since $\tilde{U}\simeq 5$\,eV in iron oxides \cite{NatMaz}, the laser strength of $eE_{0}a_{0} \simeq 10$\,eV/$\AA$ involved at the transition would burn the material. \textit{A fortiori}, reasonable strengths in iron oxides imply $Z\ll1$, so that corrections to $J^{ D}$ are too small to induce the phase transition and can only yield negligible changes, in agreement with Refs.~\cite{NatCom1, NatCom2}.

Importantly, our work shows that the presence of exchange interaction $J^{ D}$ in equilibrium is crucial to induce stronger changes in the exchange interaction $J$ with realistic laser strengths. Therefore, light control of magnetism looks more likely in \textit{p}-block materials than in \textit{d}-block transition metals. For example, the Si(111) surface doped with Pb or Sn adatoms is characterized by $t\simeq41.3$ or $43.5$\,meV, $\Delta\simeq16.7$ or $5.5$\,meV, $J^{ D}\simeq 7.3$, or $5.4$\,meV and $\tilde{U} \simeq 0.4$ or $0.5$\,eV \cite{2016arXiv160907648B}, respectively. There, a laser field of frequency $\Omega\simeq1.2$ and strength $eE_{0}a_{0}\simeq 0.75$\,eV ($a_{0}\sim4\,\AA$) would completely suppress the exchange interaction, thus inducing the anti-FM -- FM phase transition dynamically, as shown in Fig.\,\ref{fig:exchange-radius}. Note that, if the direct exchange were null in equilibrium ($J^{ D}=0$), the situation would be similar to what happens in iron oxides.

\begin{figure}[!t]
\centering
\includegraphics[width=0.55\linewidth]{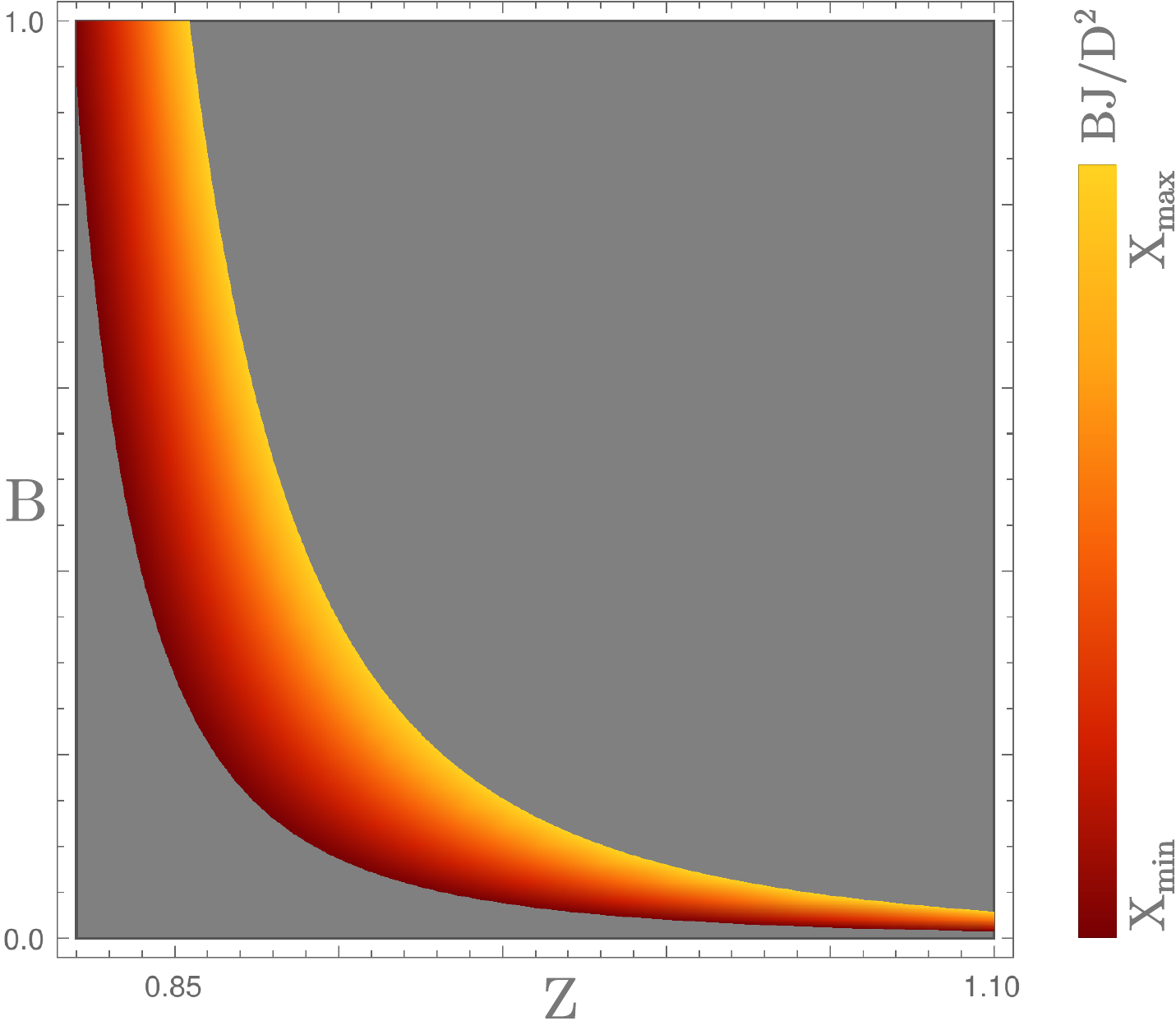}
\caption{(color online)
Stable Skyrmionic phase for the Si(111):Sn as a function of laser field amplitude $Z$ and magnetic field $\tilde{B}=B/J_{A=0}$ given in units of the initial exchange interaction $J_{A=0}$, $\Omega=4\tilde{U}$.}
\label{fig:field}
\end{figure}

The competition between exchange interaction and DMI may yield Skyrmions whose radius scales with $J/D$~\cite{BOGDANOV1994255, PSSB:PSSB2221860223, PhysRevB.82.094429}. Figure~\ref{fig:exchange-radius} shows that one can dynamically change this ratio by varying the laser strength.  Thus, it becomes possible to engineer skyrmions of arbitrary small sizes, which usually makes them easier to stabilize in experiments. Importantly, with the absence of direct exchange DMI scales in the same way as exchange interaction and the ratio $J/D$ remains almost unchanged, which is again in agreement with the Refs.~\cite{NatCom1, NatCom2}. Skyrmion stabilization can be achieved under a perpendicular magnetic field. In the case of FM Skyrmions, this occurs for magnetic fields with a strength $B$ satisfying $X_{\rm min} < \frac{BJ}{D^2} < X_{\rm max}$~\cite{PhysRevX.4.031045, PhysRevB.92.134405}. The stable Skyrmionic phase as a function of the laser field and magnetic field strengths is illustrated in Fig.~\ref{fig:field}, where the values of $X_{\rm min}$ and $X_{\rm max}$ are the ones obtained in Ref.~\onlinecite{PhysRevX.4.031045}. The left-hand side of the plot shows that the high-frequency laser can help to stabilize Skyrmions by drastically enlarging the range of suitable magnetic fields.
Anti-FM Skyrmions~\cite{AFMSk, PhysRevLett.116.147203}, however, are stabilized under high magnetic fields. For example, in Si(111):Pb, the Skyrmionic state was predicted to be stabilized under a 250 T magnetic field ~\cite{2016arXiv160907648B}. Actually, it has been shown that the strength of the magnetic field scales linearly with the exchange interaction and in particular $B\sim4J$. Therefore, shining the material with a high-frequency laser may be relevant to significantly reduce the exchange interaction and, thus, to diminish the stabilizing magnetic field down to experimentally realistic strengths.

Finally, the possibility to undergo anti-FM -- FM phase transition by varying the laser strength allows one to generate two different types of Skyrmions in one system and observe the anti-FM Skyrmionic -- FM Skyrmionic phase transition. Indeed, one can stabilize the anti-FM Skyrmions, for example, obtained in Si(111):Pb \cite{2016arXiv160907648B}, under the influence of the high-frequency light by applying a weak perpendicular magnetic field in the antiferromagnetic phase. After driving the system toward the FM phase the exchange interaction changes sign and one can then stabilize the new FM Skyrmionic structure by adjusting the magnetic field.\\

{\it{$J_1-J_{2,3}$ skyrmion model}} --
Now let us consider another interesting model that describes a frustrated magnetic system. It consists of an isotropic Heisenberg model on a triangular lattice where Skyrmions appear as a result of the competition between strong ferromagnetic NN and weak antiferromagnetic next-NN exchange interactions. In order to obtain Skyrmions, these interactions must obey special conditions that originate from the lattice structure~\cite{PhysRevLett.108.017206, leonov2015multiply, PhysRevB.93.064430, PhysRevB.93.184413}. Thus, for the $J_1-J_2$ model the exchange interactions should satisfy $-1<J_1/|J_2|<3$, whereas they should satisfy $J_1/|J_3|<4$ in the $J_1-J_3$ model. Designing such a frustrated system is of course a nontrivial problem experimentally. Nevertheless, we subsequently show that frustration can be realized by shinning an antiferromagnet with a high-frequency laser. 

We consider the single-band extended Hubbard Hamiltonian~\eqref{eq:Hamiltonian1} on a triangular lattice but with NN and next-NN hopping processes and Coulomb interactions
\begin{align}
&H = \sum_{\av{ij},\,\sigma}t^{\phantom{*}}_{1}\, c^{*}_{i\sigma}\,c^{\phantom{*}}_{j\sigma} +
\sum_{\av{\av{ij}},\,\sigma}t^{\phantom{*}}_{2,3}\,
c^{*}_{i\sigma}\,c^{\phantom{*}}_{j\sigma} +
\sum_{i} U^{\phantom{*}}_{00}\,
n^{\phantom{*}}_{i\uparrow}n^{\phantom{*}}_{i\downarrow}\\
&+\frac12\sum_{ij,\,\sigma\sigma'}\left(U^{\phantom{*}}_{\av{ij}}+
U^{\phantom{*}}_{\av{\av{ij}}}\right)
n^{\phantom{*}}_{i\sigma}n^{\phantom{*}}_{j\sigma'}
-\frac12\sum_{ij,\,\sigma\sigma'}
J^{\rm D}_{\av{ij}}\,
c^{*}_{i\sigma}c^{\phantom{*}}_{i,\sigma'}
c^{*}_{j,\sigma'}c^{\phantom{*}}_{j\sigma}. \notag
\end{align}
Using the high-frequency expansion introduced above, one can obtain an effective Hamiltonian which, for circularly polarized fields, is
\begin{align}
&{\cal H} = \sum_{\av{ij},\,\sigma}t'_{1}\, c^{*}_{i\sigma}\,c^{\phantom{*}}_{j\sigma} +
\sum_{\av{\av{ij}},\,\sigma}t'_{2,3}\,
c^{*}_{i\sigma}\,c^{\phantom{*}}_{j\sigma} +
\sum_{i} U^{\phantom{*}}_{00}\,
n^{\phantom{*}}_{i\uparrow}n^{\phantom{*}}_{i\downarrow}\\
&+\frac12\sum_{ij,\,\sigma\sigma'}\left(U^{\phantom{*}}_{\av{ij}}+
U^{\phantom{*}}_{\av{\av{ij}}}\right)
n^{\phantom{*}}_{i\sigma}n^{\phantom{*}}_{j\sigma'}
-\frac12\sum_{ij,\,\sigma\sigma'}
J'^{\rm D}_{\av{ij}}\,
c^{*}_{i\sigma}c^{\phantom{*}}_{i,\sigma'}
c^{*}_{j,\sigma'}c^{\phantom{*}}_{j\sigma}, \notag
\end{align}
where the renormalized hopping amplitudes are $t'_1 = t_1{\cal J}_0(Z)$, $t'_2 = t_2{\cal J}_0(\sqrt{3}Z)$, and $t'_3 = t_3{\cal J}_0(2Z)$. The explicit expression of the renormalized exchange interaction is detailed in Ref.~\cite{SM}.

\begin{figure}[!t]
\centering
\includegraphics[width=1\linewidth]{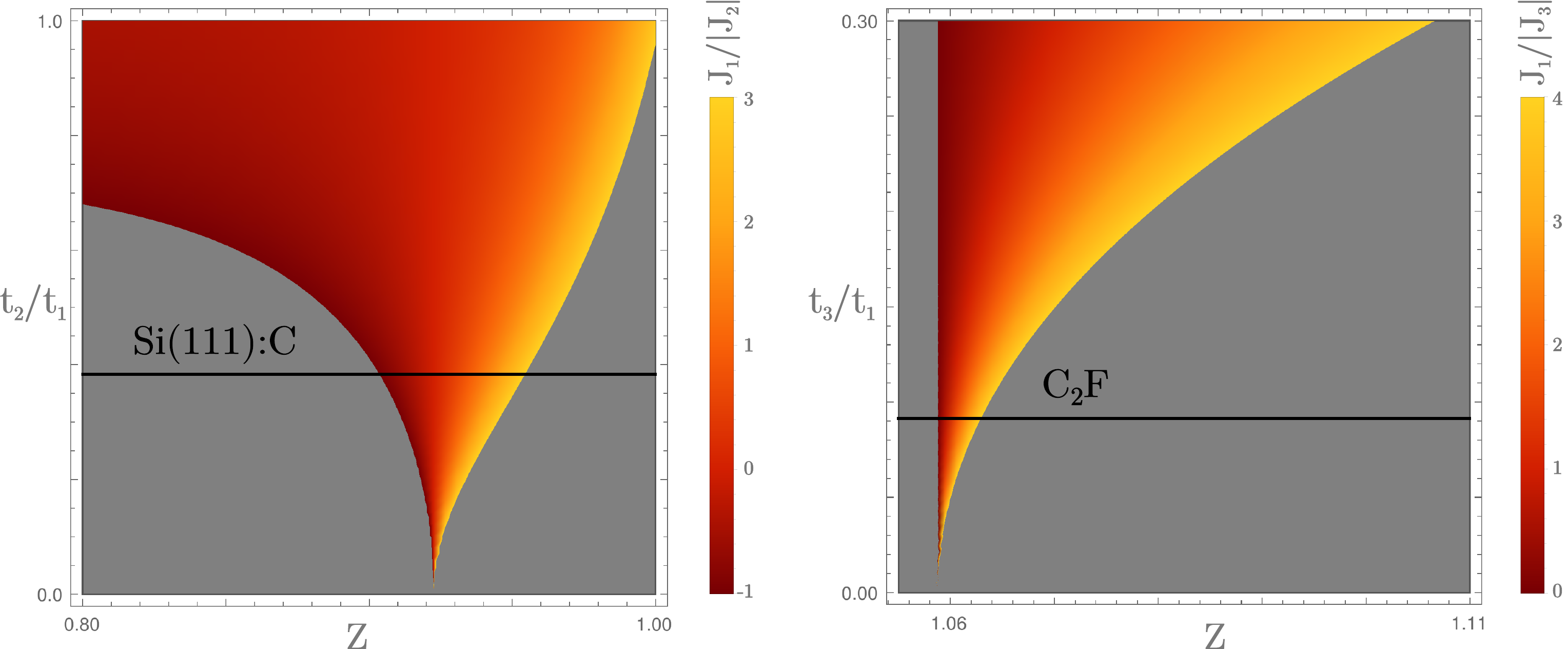}
\caption{(color online) Hopping amplitudes $t_{2(3)}/t_1$ as the function of the amplitude $Z$ of the laser field for the values $J_1/|J_{2(3)}|$ that correspond to the Skyrmionic phase. Frequency of the laser field is $\Omega=3\tilde{U}$.}
\label{fig:t1-t3}
\end{figure}

When the system lies in the strong interaction regime, one can write an effective Heisenberg model,
\begin{align}
&H_{\rm spin} = -\sum_{\av{ij}}J_{1}\,\hat{\bf S}_{i}\,\hat{\bf S}_{j} -
\sum_{\av{\av{ij}}}J_{2,3}\,\hat{\bf S}_{i}\,\hat{\bf S}_{j}
\end{align}
with NN exchange interaction $J_{1} = J'^{ D}_{\av{ij}}- 2t'^2_1/\tilde{U}_{\av{ij}}$ and next-NN exchange interaction $J_{2,3} = J'^{ D}_{\av{\av{ij}}}-2t'^2_{2,3}/\tilde{U}_{\av{\av{ij}}}$. Top right panel in Fig.~\ref{fig:exchange-radius} shows that, for vanishing laser fields, the system lies in the antiferromagnetic phase. When turning on the laser field and increasing its strength, the system undergoes a transition toward a ferromagnetic phase. Importantly, the nearest-neighbor and the next-NN exchange interactions, namely, $J_{1}$ and $J_{2,3}$, become ferromagnetic for different values of the field, meaning that one can engineer a frustrated magnet. Here we took the ``unrealistic'' case of $t_1=t_2=t_3$ just to make the anti-FM -- FM transition more visible in the figure. Fig.~\ref{fig:t1-t3} shows the phase diagram based on conditions $-1<J_1/|J_2|<3$ and $J_1/|J_3|<4$, as a function of $t_{2,3}/t_{1}$ and laser strength $Z$. Thus, the initial antiferromagnet may be dynamically driven toward the frustrated magnetic system predicted in Ref.~\onlinecite{PhysRevLett.108.017206} with suitable values of anti-FM and FM exchange interactions to obtain Skyrmions. In the case of Si(111) with C adatoms, it is estimated that $t_{01}\simeq35.1$\,meV, $t_{02}\simeq-13.5$\,meV, $J^{ D}\simeq1.67$\,meV, $\tilde{U}_{01}\simeq0.9$\,eV, $\tilde{U}_{02}\simeq1.1$\,eV \cite{2016arXiv160907648B}, so that fields with frequency $\Omega\simeq 2.7$\,eV and amplitudes $eE_{0}a_{0} \simeq \Omega$ would induce suitable values of $J_{1,2}$ to obtain Skyrmions, according to the left panel in Fig.~\ref{fig:t1-t3}. Similar effects are predicted in C$_2$F, where $t_{01}\simeq-232.8$\,meV, $t_{03}\simeq-21.3$\,meV, $J^{ D}\simeq 20$\,meV and $\tilde{U}_{01} \simeq 2.7$\,eV, $\tilde{U}_{03} \simeq~ 3.7$\,eV~\cite{Rudenko} (see right panel of Fig.~\ref{fig:t1-t3}).

So far we have only considered the case of a circular polarization. For example, in the case of the square lattice under the influence of the noncircular polarized fields, the hopping amplitude and spin-orbit coupling vector are renormalized by the Bessel functions ${\cal J}_0(eA_{x (y)}a_{0})$, where the labels $x (y)$ correspond to the direction of the vector that connects two lattice sites. This allows us to change DMI and the Skyrmion radius $J/D$ in an anisotropic way. This case was recently investigated in Ref.~\onlinecite{shibata2015large}, where DMI are tuned by strain forces, which changes the Skyrmion shape from circular to elliptic.\\

To summarize, we have reported the possibility to dynamically control the intrinsic magnetic interactions of two-dimensional materials. This can induce drastic changes in the anti-FM exchange interaction that can even be switched to FM, provided the direct exchange interaction in equilibrium is non-negligible. Additionally, the high-frequency laser field also renormalizes the DMI, so that Skyrmion features such that their radius can be tuned too, thus making them easier to stabilize under perpendicular magnetic fields. Besides, it has been shown that a high-frequency laser field can also induce dynamical frustration in antiferromagnetic triangular lattices, where the degree of frustration can be tuned suitably to experience Skyrmions.
Importantly, the dynamical effects we have discussed within the high-frequency limit rely on laser strengths and frequencies that remain reasonable for realizations in solid state physics. In particular, we expect them to be relevant when irradiating \textit{sp} and \textit{p} materials like C$_2$F and Si(111):\{C, Sn, Pb\}.\\

\acknowledgements
The authors thank A. N. Rudenko, V. V. Mazurenko, T. Kuwahara and A. Kimel for fruitful discussions and comments. This work was supported by NWO via Spinoza Prize and by ERC Advanced Grant 338957 FEMTO/NANO. Also, E. A. S. and M. I. K. acknowledge the Stichting voor Fundamenteel Onderzoek der Materie (FOM), which is financially supported by the Nederlandse Organisatie
voor Wetenschappelijk Onderzoek (NWO).

\bibliographystyle{apsrev4-1}
\bibliography{Skyrmions_3}

\onecolumngrid
\vspace{1cm}
\section{Supplemental Material for ``Dynamical and reversible control of topological spin textures''
}
\begin{center}
E. A. Stepanov$^1$, C. Dutreix$^{1,2}$, M. I. Katsnelson$^1$ \\
{\it\small $^1$Radboud University, Institute for Molecules and Materials, Heyendaalseweg 135, 6525AJ Nijmegen, Netherlands\\
$^2$Univ Lyon, Ens de Lyon, Univ Claude Bernard, CNRS, Laboratoire de Physique, F-69342 Lyon, France}
\end{center}

\subsection{Fourier transform of the kinetic part of the Hubbard-like Hamiltonian with DMI}
The Fourier transform of the time-dependent part of the Hamiltonian that accounts for the effects of the high-frequency laser field can be obtained with the use of the followgin well-known relations which arise from the definition of the Bessel function of the $m$-th order ${\cal J}_{m}(Z)$:
\begin{align}
&\int_{-\pi}^{+\pi}\frac{dt}{2\pi}
e^{-iZ_{y}\sin(t-\phi)}e^{imt} =
\int_{-\pi}^{+\pi}\frac{dt}{2\pi}
e^{-iZ_{y}\sin{}t'}e^{im(t'+\phi)} =
e^{im\phi}{\cal J}_{m}(Z_{y}),\notag\\
&\int_{-\pi}^{+\pi}\frac{dt}{2\pi}
e^{-iZ_{x}\cos{}t-iZ_{y}\sin(t-\phi)}e^{imt} =
e^{im\theta}{\cal J}_{m}\sqrt{Z^2_{x}+Z^2_{y} - 2Z_{x}Z_{y}\sin\phi},\notag
\end{align}
where $\cot\theta=\frac{Z_{y}\cos\phi}{Z_{y}\sin\phi-Z_{x}}$. This results in the following expression for the Fourier transform of the hopping amplitude and the spin-orbit coupling
\begin{align}
\varepsilon_{\kv,m} &=
2t{\cal J}_{m}(Z)\left[\cos(k_{x}-m\pi/2)
+\cos(k_{x}/2+k_{y}\sqrt{3}/2)\,
e^{im5\pi/6}+\cos(k_{x}/2-k_{y}\sqrt{3}/2)\,
e^{im\pi/6}\right], \notag\\
f^{x}_{\kv,m} &=\sqrt{3}i\Delta{}{\cal J}_{m}(Z)\left[\sin(k_{x}/2-k_{y}\sqrt{3}/2)e^{im\pi/6}-\sin(k_{x}/2+k_{y}\sqrt{3}/2)e^{im5\pi/6}\right], \notag\\
f^{y}_{\kv,m} &= i\Delta{}{\cal J}_{m}(Z) \left[ 2\sin(k_{x}-m\pi/2) + \sin(k_{x}/2-k_{y}\sqrt{3}/2)e^{im\pi/6}+\sin(k_{x}/2+k_{y}\sqrt{3}/2)e^{im5\pi/6}\right], 
\end{align}
where we consider the case of the circularly polarized light ($Z = eA_{x}a_0=eA_{y}a_0$, $\phi=0$) for simplicity. 
Then, the full Hamiltonian in the frequency space can be written as
\begin{align}
H_{m} =& \sum_{\kv,\,\sigma\sigma'} c^{*}_{\kv\sigma}\left(\varepsilon^{\phantom{*}}_{\kv,m}
\delta^{\phantom{*}}_{\sigma\sigma'} + i\boldsymbol{f}^{\phantom{*}}_{\kv,m}
\boldsymbol{\sigma}^{\phantom{*}}_{\sigma\sigma'}
\right)c^{\phantom{*}}_{\kv\sigma'} + V\delta^{\phantom{*}}_{m0},
\label{eq:Hamfr}
\end{align}
where the time-independent interaction term transforms as $V\delta^{\phantom{*}}_{m0}$.

\subsection{High-frequency description}
As it was mentioned in the main text, the time-periodic Hamiltonian obey the time-dependent Schr\"odinger equation
\begin{align}
i \partial_\tau \Psi(\lambda, \tau) =~\frac{1}{\Omega} H(\tau) \Psi (\lambda, \tau),
\end{align}
where $\tau=\Omega t$. One can introduce a dimensionless parameter $\lambda=\delta E/ \Omega$ which compares a certain energy scale $\delta E$ to the typical field energy. For simplicity we chose $\delta E$ as the largest characteristic energy involved in the initial Hamiltonian among $t_{ij}$, $U_{ij}$ and $J^{\rm D}_{ij}$, so that no resonant processes with the applied laser field will occur. Then, the Schr\"odinger equation can be rewritten as
\begin{align}
i \partial_\tau \Psi(\lambda, \tau) =~\lambda \overline{H}(\tau)\Psi (\lambda, \tau),
\end{align}
where the Hamiltonian is renormalized on the same energy scale as
\begin{align}
\frac{1}{\Omega}H(\tau)=\lambda{}\frac{H(\tau)}{\delta{}E}=\lambda{}\overline{H}(\tau).
\end{align}
In order to obtain the effective Hamiltonian of our model we look for a unitary transformation defined as
\begin{align}
\Psi (\lambda, \tau) = \exp\{-i\Delta(\tau)\}\,\psi (\lambda, \tau),
\end{align}
which would remove the time dependence of the Hamiltonian. By construction we also impose
\begin{align}
\Delta(\tau) =~ \sum_{n=1}^{+\infty} \lambda^{n} \Delta_{n}(\tau)
\end{align}
with $\Delta_{n}(\tau)$ a $2\pi$ periodic function that averages at zero. Such a transformation leads to
\begin{align}
i\partial_\tau \psi(\lambda, \tau) = \frac{1}{\Omega}{\cal H} \psi (\lambda, \tau) = \lambda \overline{\cal H}\psi (\lambda, \tau)
\end{align}
with effective Hamiltonian
\begin{align}\label{Effective Hamiltonian}
\overline{\cal H}=e^{i\Delta(\tau)} \overline{H}(\tau)\,e^{-i\Delta(\tau)} -i\lambda^{-1} e^{i\Delta(\tau)} \partial_{\tau} e^{-i\Delta(\tau)} ~,
\end{align}
or equivalently
\begin{align}
{\cal H}=e^{i\Delta(\tau)}H(\tau)\,e^{-i\Delta(\tau)} -i\Omega e^{i\Delta(\tau)} \partial_{\tau} e^{-i\Delta(\tau)} ~.
\end{align}
The partial time-derivative satisfies the following relation
\begin{align}\label{Exponential Operator Derivative}
\partial_{\tau} e^{-i\Delta(\tau)} &= \sum_{n=0}^{\infty} \frac{ \big\{ \big(\!-i\Delta(\tau) \big)^{n}, -\,i \partial_{\tau}\Delta(\tau) \big\}}{(n+1)!}~ e^{-i\Delta(\tau)} ~,
\end{align}
where the repeated commutator is defined for two operators $X$ and $Y$ by $\{1, Y\}=Y$ and $\{X^{n}, Y\}=~[X, \{ X^{n-1}, Y\}]$. The square brackets denote the usual commutator. Then, one can write
\begin{align}
\overline{\cal H}= e^{i\Delta(\tau)} \left[\overline{H}(\tau) -i\lambda^{-1} \sum_{n=0}^{\infty} \frac{ \big\{ \big(\!-i\Delta(\tau) \big)^{n}, -\,i \partial_{\tau}\Delta(\tau) \big\}}{(n+1)!} \right] e^{-i\Delta(\tau)} ~,
\end{align}
Using the series representation
\begin{align} \overline{\cal H}=\sum_{n=0}^{\infty}\lambda^{n}
\tilde{H}_{n},
\end{align}
together with Eqs. \eqref{Exponential Operator Derivative} and \eqref{Effective Hamiltonian}, one can then determine operators $\tilde{H}_{n}$ and $\Delta_{n}$ iteratively in all orders in $\lambda$. Here, we restrict ourselves to the case of the high-frequency laser field, which allows us to consider the effective Hamiltonian representation up to the second order correction in $\lambda$: $\overline{\cal H} = \tilde{H}_{0} + \lambda\tilde{H}_{1} + \lambda^{2}\tilde{H}_{2}$. These effective time-independent Hamiltonians describe the stroboscopic dynamics of the system, whereas its evolution between two stroboscopic times is encoded into the time-dependent function $\Delta_{n}(\tau)$~\cite{PhysRevB.93.241404}.

As it was showed in the Ref.~\cite{PhysRevB.93.241404}, the first term in this representation is given by the time-average $\tilde{H}_0=\av{\overline{H}(\tau)}=\overline{H}_{0}$, where
\begin{align}
\overline{H}_{m}=\int_{-\pi}^{+\pi} \frac{d\tau}{2\pi}~ e^{im\tau} \overline{H}(\tau).
\end{align}
Since in the initial problem the interaction term was time-independent, the time-averaging procedure changes only the single-particle terms of the Hamiltonian $\overline{H}_{m}$, which results in the renormalization of the hopping amplitude and spin-orbit coupling with respect to the time-independent problem as
\begin{align}
{\cal H}\simeq\tilde{H}_{0}\,\delta{}E =
\sum_{\av{ij},\,\sigma\sigma'}c^{*}_{i\sigma}
\left(t'\,\delta^{\phantom{*}}_{\sigma\sigma'} + 
i\boldsymbol{\Delta}'_{ij}\,
\boldsymbol{\sigma}^{\phantom{*}}_{\sigma\sigma'}
\right)c^{\phantom{*}}_{j\sigma'}
+ \sum_{i}U^{\phantom{*}}_{00}\,
n^{\phantom{*}}_{i\uparrow}n^{\phantom{*}}_{i\downarrow} 
+\frac12\sum_{\av{ij},\sigma\sigma'}U^{\phantom{*}}_{\av{ij}}\,
n^{\phantom{*}}_{i\sigma}n^{\phantom{*}}_{j\sigma'}
-\frac12\sum_{\av{ij},\,\sigma\sigma'}
J^{\rm D}_{\av{ij}}\,
c^{*}_{i\sigma}c^{\phantom{*}}_{i\sigma'}
c^{*}_{j\sigma'}c^{\phantom{*}}_{j\sigma}, \notag
\end{align}
where
\begin{align}
t' = t\,{\cal J}_0(Z) ~~ \text{and} ~~
\Delta' = \Delta\,{\cal J}_0(Z).
\end{align}

\subsection{First-order correction $\tilde{H}_1$ to the time-averaged Hamiltonian in the high-frequency expansion}
The first-order in $\lambda$ term in the effective Hamiltonian is given by the following equation
\begin{align}
\tilde{H}_1 = -\sum_{m>0}\frac{[\overline{H}_{m},\overline{H}_{-m}]}{m}.
\end{align}
Since $m\neq0$ does not contribute to the first-order term $\tilde{H}_{1}$, one can rewrite the time dependent part of the Hamiltonian as follows
\begin{align}
\overline{H}_{m>0} =&\sum_{\kv}
\overline{\varepsilon}^{\phantom{*}}_{\kv,m}\left(c^{*}_{\kv\uparrow}c^{\phantom{*}}_{\kv\uparrow} +
c^{*}_{\kv\downarrow}c^{\phantom{*}}_{\kv\downarrow}\right)+
i\sum_{\kv}\overline{f}^{x}_{\kv,m}\left(c^{*}_{\kv\uparrow}c^{\phantom{*}}_{\kv\downarrow} +
c^{*}_{\kv\downarrow}c^{\phantom{*}}_{\kv\uparrow}\right) +
\sum_{\kv}\overline{f}^{y}_{\kv,m}\left(c^{*}_{\kv\uparrow}c^{\phantom{*}}_{\kv\downarrow} -
c^{*}_{\kv\downarrow}c^{\phantom{*}}_{\kv\uparrow}\right),
\end{align}
where we use the same notations for the renormalized variables $\overline{\varepsilon}_{\kv}=\varepsilon_{\kv}/\delta{}E$ and $\overline{f}^{x,y}_{\kv}=f^{x,y}_{\kv}/\delta{}E$.\\
Let us compute the general commutator $[\overline{H}_{m},\overline{H}_{n}]$ ($m,n>0$) that in our case splits into the three different terms
\begin{subequations}
\begin{align}
\overline{\varepsilon}^{\phantom{*}}_{\kv_1,m}\overline{\varepsilon}^{\phantom{*}}_{\kv_2,n}
&\left[\left(c^{*}_{\kv_1\uparrow}c^{\phantom{*}}_{\kv_1\uparrow} +
c^{*}_{\kv_1\downarrow}c^{\phantom{*}}_{\kv_1\downarrow}\right),
\left(c^{*}_{\kv_2\uparrow}c^{\phantom{*}}_{\kv_2\uparrow} +
c^{*}_{\kv_2\downarrow}c^{\phantom{*}}_{\kv_2\downarrow}\right)\right],\label{app:H2a}\\
\overline{\varepsilon}^{\phantom{*}}_{\kv_1,m}\overline{f}^{i}_{\kv_2,n}
&\left[\left(c^{*}_{\kv_1\uparrow}c^{\phantom{*}}_{\kv_1\uparrow} +
c^{*}_{\kv_1\downarrow}c^{\phantom{*}}_{\kv_1\downarrow}\right),
\left(c^{*}_{\kv_2\uparrow}c^{\phantom{*}}_{\kv_2\downarrow} \mp
c^{*}_{\kv_2\downarrow}c^{\phantom{*}}_{\kv_2\uparrow}\right)\right],\label{app:H2b}\\
\overline{f}^{i}_{\kv_1,m}\overline{f}^{j}_{\kv_2,n}
&\left[\left(c^{*}_{\kv_1\uparrow}c^{\phantom{*}}_{\kv_1\downarrow} \mp
c^{*}_{\kv_1\downarrow}c^{\phantom{*}}_{\kv_1\uparrow}\right),
\left(c^{*}_{\kv_2\uparrow}c^{\phantom{*}}_{\kv_2\downarrow} \mp
c^{*}_{\kv_2\downarrow}c^{\phantom{*}}_{\kv_2\uparrow}\right)\right].\label{app:H2c}
\end{align}
\end{subequations}
Let us start from Eq.~\eqref{app:H2a}. Using the commutation relations one can obtain that
\begin{align}
c^{*}_{\kv_1\uparrow}c^{\phantom{*}}_{\kv_1\uparrow}
c^{*}_{\kv_2\uparrow}c^{\phantom{*}}_{\kv_2\uparrow}
&= -\,c^{*}_{\kv_1\uparrow}c^{*}_{\kv_2\uparrow}
c^{\phantom{*}}_{\kv_1\uparrow}c^{\phantom{*}}_{\kv_2\uparrow} +
\delta^{\phantom{*}}_{\kv_1,\kv_2} c^{*}_{\kv_1\uparrow}c^{\phantom{*}}_{\kv_2\uparrow}
= -\,c^{*}_{\kv_2\uparrow}c^{*}_{\kv_1\uparrow}
c^{\phantom{*}}_{\kv_2\uparrow}c^{\phantom{*}}_{\kv_1\uparrow} +
\delta^{\phantom{*}}_{\kv_1,\kv_2} c^{*}_{\kv_1\uparrow}c^{\phantom{*}}_{\kv_2\uparrow} \notag\\
&=\phantom{-}\,c^{*}_{\kv_2\uparrow}c^{\phantom{*}}_{\kv_2\uparrow} c^{*}_{\kv_1\uparrow}c^{\phantom{*}}_{\kv_1\uparrow} -
\delta^{\phantom{*}}_{\kv_1,\kv_2}
c^{*}_{\kv_2\uparrow}c^{\phantom{*}}_{\kv_1\uparrow}
+ \delta^{\phantom{*}}_{\kv_1,\kv_2} c^{*}_{\kv_1\uparrow}c^{\phantom{*}}_{\kv_2\uparrow} =c^{*}_{\kv_2\uparrow}c^{\phantom{*}}_{\kv_2\uparrow} c^{*}_{\kv_1\uparrow}c^{\phantom{*}}_{\kv_1\uparrow}, \notag\\
c^{*}_{\kv_1\downarrow}c^{\phantom{*}}_{\kv_1\downarrow}
c^{*}_{\kv_2\uparrow}c^{\phantom{*}}_{\kv_2\uparrow}
&=\phantom{-}\,c^{*}_{\kv_2\uparrow}c^{\phantom{*}}_{\kv_2\uparrow}
c^{*}_{\kv_1\downarrow}c^{\phantom{*}}_{\kv_1\downarrow}.
\end{align}
Making the same transformations with other remaining terms, one can see that commutator in Eq.~\eqref{app:H2a} is equal to zero.\\
The result of Eq.~\eqref{app:H2b} can be obtained in the same style.
\begin{align}
c^{*}_{\kv_1\uparrow}c^{\phantom{*}}_{\kv_1\uparrow}
c^{*}_{\kv_2\uparrow}c^{\phantom{*}}_{\kv_2\downarrow} +
c^{*}_{\kv_1\downarrow}c^{\phantom{*}}_{\kv_1\downarrow}
c^{*}_{\kv_2\uparrow}c^{\phantom{*}}_{\kv_2\downarrow}
&= -\,c^{*}_{\kv_1\uparrow}c^{*}_{\kv_2\uparrow}
c^{\phantom{*}}_{\kv_1\uparrow}c^{\phantom{*}}_{\kv_2\downarrow} +
\delta^{\phantom{*}}_{\kv_1\kv_2}c^{*}_{\kv_1\uparrow}c^{\phantom{*}}_{\kv_2\downarrow} -
c^{*}_{\kv_1\downarrow}c^{*}_{\kv_2\uparrow}
c^{\phantom{*}}_{\kv_1\downarrow}c^{\phantom{*}}_{\kv_2\downarrow} \\
&= -\,c^{*}_{\kv_2\uparrow}c^{*}_{\kv_1\uparrow}
c^{\phantom{*}}_{\kv_2\downarrow}c^{\phantom{*}}_{\kv_1\uparrow} +
\delta^{\phantom{*}}_{\kv_1\kv_2}c^{*}_{\kv_1\uparrow}c^{\phantom{*}}_{\kv_2\downarrow} -
c^{*}_{\kv_2\uparrow}c^{*}_{\kv_1\downarrow}
c^{\phantom{*}}_{\kv_2\downarrow}c^{\phantom{*}}_{\kv_1\downarrow} \notag\\
&=\phantom{-}\,c^{*}_{\kv_2\uparrow}c^{\phantom{*}}_{\kv_2\downarrow}
c^{*}_{\kv_1\uparrow}c^{\phantom{*}}_{\kv_1\uparrow} +
\delta^{\phantom{*}}_{\kv_1\kv_2}c^{*}_{\kv_1\uparrow}c^{\phantom{*}}_{\kv_2\downarrow} +
c^{*}_{\kv_2\uparrow}c^{\phantom{*}}_{\kv_2\downarrow}
c^{*}_{\kv_1\downarrow}c^{\phantom{*}}_{\kv_1\downarrow} -
\delta^{\phantom{*}}_{\kv_1\kv_2}c^{*}_{\kv_2\uparrow}c^{\phantom{*}}_{\kv_1\downarrow} \notag\\
&=\phantom{-}\,c^{*}_{\kv_2\uparrow}c^{\phantom{*}}_{\kv_2\downarrow}
c^{*}_{\kv_1\uparrow}c^{\phantom{*}}_{\kv_1\uparrow} +
c^{*}_{\kv_2\uparrow}c^{\phantom{*}}_{\kv_2\downarrow}
c^{*}_{\kv_1\downarrow}c^{\phantom{*}}_{\kv_1\downarrow}. \notag
\end{align}
Obtaining the second term with the similar transformations, one obtains that commutator in Eq.~\eqref{app:H2b} is also equal to zero.\\
The last commutator given by Eq.~\eqref{app:H2c} is zero as well. Indeed, since
\begin{align}
c^{*}_{\kv_1\downarrow}c^{\phantom{*}}_{\kv_1\uparrow}
c^{*}_{\kv_2\uparrow}c^{\phantom{*}}_{\kv_2\downarrow} +
c^{*}_{\kv_1\uparrow}c^{\phantom{*}}_{\kv_1\downarrow}
c^{*}_{\kv_2\downarrow}c^{\phantom{*}}_{\kv_2\uparrow}
&= -\,c^{*}_{\kv_1\downarrow}c^{*}_{\kv_2\uparrow}
c^{\phantom{*}}_{\kv_1\uparrow}c^{\phantom{*}}_{\kv_2\downarrow} +
\delta^{\phantom{*}}_{\kv_1\kv_2}
c^{*}_{\kv_1\downarrow}c^{\phantom{*}}_{\kv_2\downarrow}
-c^{*}_{\kv_1\uparrow}c^{*}_{\kv_2\downarrow}
c^{\phantom{*}}_{\kv_1\downarrow}c^{\phantom{*}}_{\kv_2\uparrow} +
\delta^{\phantom{*}}_{\kv_1\kv_2}
c^{*}_{\kv_1\uparrow}c^{\phantom{*}}_{\kv_2\uparrow} \\
&= -\,c^{*}_{\kv_2\uparrow}c^{*}_{\kv_1\downarrow}
c^{\phantom{*}}_{\kv_2\downarrow}c^{\phantom{*}}_{\kv_1\uparrow} +
\delta^{\phantom{*}}_{\kv_1\kv_2}
c^{*}_{\kv_1\downarrow}c^{\phantom{*}}_{\kv_2\downarrow}
-c^{*}_{\kv_2\downarrow}c^{*}_{\kv_1\uparrow}
c^{\phantom{*}}_{\kv_2\uparrow}c^{\phantom{*}}_{\kv_1\downarrow} +
\delta^{\phantom{*}}_{\kv_1\kv_2}
c^{*}_{\kv_1\uparrow}c^{\phantom{*}}_{\kv_2\uparrow} \notag\\
&=\phantom{-}\,c^{*}_{\kv_2\uparrow}c^{\phantom{*}}_{\kv_2\downarrow}
c^{*}_{\kv_1\downarrow}c^{\phantom{*}}_{\kv_1\uparrow} -
\delta^{\phantom{*}}_{\kv_1\kv_2}
c^{*}_{\kv_2\uparrow}c^{\phantom{*}}_{\kv_1\uparrow} +
\delta^{\phantom{*}}_{\kv_1\kv_2}
c^{*}_{\kv_1\downarrow}c^{\phantom{*}}_{\kv_2\downarrow} +
c^{*}_{\kv_2\downarrow}c^{\phantom{*}}_{\kv_2\uparrow}
c^{*}_{\kv_1\uparrow}c^{\phantom{*}}_{\kv_1\downarrow} -
\delta^{\phantom{*}}_{\kv_1\kv_2}
c^{*}_{\kv_2\downarrow}c^{\phantom{*}}_{\kv_1\downarrow} +
\delta^{\phantom{*}}_{\kv_1\kv_2}
c^{*}_{\kv_1\uparrow}c^{\phantom{*}}_{\kv_2\uparrow} \notag\\
&=\phantom{-}\,c^{*}_{\kv_2\uparrow}c^{\phantom{*}}_{\kv_2\downarrow}
c^{*}_{\kv_1\downarrow}c^{\phantom{*}}_{\kv_1\uparrow} +
c^{*}_{\kv_2\downarrow}c^{\phantom{*}}_{\kv_2\uparrow}
c^{*}_{\kv_1\uparrow}c^{\phantom{*}}_{\kv_1\downarrow}\notag,\\
c^{*}_{\kv_1\uparrow}c^{\phantom{*}}_{\kv_1\downarrow}
c^{*}_{\kv_2\uparrow}c^{\phantom{*}}_{\kv_2\downarrow}
&=\phantom{-}\,
c^{*}_{\kv_2\uparrow}c^{\phantom{*}}_{\kv_2\downarrow}
c^{*}_{\kv_1\uparrow}c^{\phantom{*}}_{\kv_1\downarrow},
\end{align}
the commutator in Eq.~\eqref{app:H2c} is equal to zero.
So, if $[\overline{H}_{m}, \overline{H}_{n}]=0$ for all $m,n>0$, the first-order correction term $\tilde{H}_1$ of the effective Hamiltonian is identically zero.

\subsection{Second-order term $\tilde{H}_2$ of the Hamiltonian in the high-frequency expansion}
Since $[\overline{H}_{m}, \overline{H}_{n}]=0$ as it was shown above, the second-order term $\tilde{H}_{2}$ of the effective Hamiltonian can be simplified as
\begin{align}
\tilde{H}_2 = \sum_{m>0}\frac{[[\overline{H}_{m},\overline{H}_0],\overline{H}_{-m}]}{m^2} =
\sum_{m>0}\frac{[[\overline{H}_{m},\overline{V}],\overline{H}_{-m}]}{m^2}.
\label{eq:H3}
\end{align}
It should be mentioned, that we consider the high-frequency description of the effective Hamiltonian until the second order in $\lambda$. In general, one can stop at the first-order term $\tilde{H}_{1}$, because the second order term $\tilde{H}_2$ describes only the corrections to the interactions, that will be $\lambda^2$ times smaller than the interactions already described by the time averaged term $\tilde{H}_0$. Therefore, they are negligibly small and \textit{a priori} not important for the magnetic properties of the system. Nevertheless, according to the Ref.~\cite{Itin2}, the high-frequency field generates new types of interaction terms that were not present in the initial problem. In particular, there is one important correction, namely $J^{\rm D}_{\rm ind}$, to the direct exchange interaction term $J^{\rm D}$ that comes from the Coulomb interaction and appears exactly in the second-order correction $\tilde{H}_2$ in the effective Hamiltonian. After taking this contribution into account, the next-order terms are negligible because they will again be $\lambda^2$ times smaller than the already accounted terms. This allows us to choose $\lambda^3\ll1$  and stop at the second-order in $\lambda$ term $\tilde{H}_2$ while constructing the effective Hamiltonian of considered problem. Also, since $J^{\rm D}\ll{}U$ in every magnetic system, and because they are renormalized by the same Bessel functions ${\cal J}_{m}(Z)$, it allows us to simplify the commutation in Eq.~\eqref{eq:H3} as
\begin{align}
\tilde{H}_2 = \sum_{\substack{\qv \\ m>0}}
\sum_{\substack{\kv_1, \kv_2, \kv_3, \kv_4 \\ \sigma_1, \sigma_2, \sigma_3, \sigma_4}}
\frac{\left[\left[\overline{H}_{m}(\kv_1,\sigma_1),
\frac12\overline{U}^{\phantom{*}}_{\qv}c^{*}_{\kv_2,\sigma_2}c^{\phantom{*}}_{\kv_2+\qv,\sigma_2}
c^{*}_{\kv_3,\sigma_3}c^{\phantom{*}}_{\kv_3-\qv,\sigma_3}
\right],
\overline{H}_{-m}(\kv_4,\sigma_4)\right]}{m^2}.
\label{eq:H3s}
\end{align}
Although in the many systems the spin-orbit coupling is much smaller than the hopping amplitude $\Delta\ll{}t$, we will not neglect it in the Eq.\eqref{eq:H3s} and will show that contrary to the case of exchange interaction, the effective DMI interaction is not induced by the high-frequency laser field.

Let us study the general commutator
$\Big[c^{*}_{\kv_1,\sigma_1}c^{\phantom{*}}_{\kv_1,\sigma_2},
c^{*}_{\kv_2,\sigma_3}c^{\phantom{*}}_{\kv_2+\qv,\sigma_4}
c^{*}_{\kv_3,\sigma_5}c^{\phantom{*}}_{\kv_3-\qv,\sigma_6}\Big]$
\begin{align}
c^{*}_{\kv_1,\sigma_1}c^{\phantom{*}}_{\kv_1,\sigma_2}
c^{*}_{\kv_2,\sigma_3}c^{\phantom{*}}_{\kv_2+\qv,\sigma_4}
c^{*}_{\kv_3,\sigma_5}c^{\phantom{*}}_{\kv_3-\qv,\sigma_6}
&= - \,
c^{*}_{\kv_1,\sigma_1}c^{*}_{\kv_2,\sigma_3}
c^{\phantom{*}}_{\kv_1,\sigma_2}c^{\phantom{*}}_{\kv_2+\qv,\sigma_4}
c^{*}_{\kv_3,\sigma_5}c^{\phantom{*}}_{\kv_3-\qv,\sigma_6} +
\delta^{\phantom{*}}_{\kv_1,\kv_2}\delta^{\phantom{*}}_{\sigma_2,\sigma_3}
c^{*}_{\kv_1,\sigma_1}c^{\phantom{*}}_{\kv_2+\qv,\sigma_4}
c^{*}_{\kv_3,\sigma_5}c^{\phantom{*}}_{\kv_3-\qv,\sigma_6} \notag\\
&= - \,
c^{*}_{\kv_2,\sigma_3}c^{*}_{\kv_1,\sigma_1}
c^{\phantom{*}}_{\kv_2+\qv,\sigma_4}c^{\phantom{*}}_{\kv_1,\sigma_2}
c^{*}_{\kv_3,\sigma_5}c^{\phantom{*}}_{\kv_3-\qv,\sigma_6} +
\delta^{\phantom{*}}_{\kv_1,\kv_2}\delta^{\phantom{*}}_{\sigma_2,\sigma_3}
c^{*}_{\kv_1,\sigma_1}c^{\phantom{*}}_{\kv_2+\qv,\sigma_4}
c^{*}_{\kv_3,\sigma_5}c^{\phantom{*}}_{\kv_3-\qv,\sigma_6} \notag\\
&= \phantom{-}\,
c^{*}_{\kv_2,\sigma_3}c^{\phantom{*}}_{\kv_2+\qv,\sigma_4}
c^{*}_{\kv_1,\sigma_1}c^{\phantom{*}}_{\kv_1,\sigma_2}
c^{*}_{\kv_3,\sigma_5}c^{\phantom{*}}_{\kv_3-\qv,\sigma_6} +
\delta^{\phantom{*}}_{\kv_1,\kv_2}\delta^{\phantom{*}}_{\sigma_2,\sigma_3}
c^{*}_{\kv_1,\sigma_1}c^{\phantom{*}}_{\kv_2+\qv,\sigma_4}
c^{*}_{\kv_3,\sigma_5}c^{\phantom{*}}_{\kv_3-\qv,\sigma_6} \notag\\
&\hspace{5.8cm}
-\delta^{\phantom{*}}_{\kv_1,\kv_2+\qv}\delta^{\phantom{*}}_{\sigma_1,\sigma_4}
c^{*}_{\kv_2,\sigma_3}c^{\phantom{*}}_{\kv_1,\sigma_2}
c^{*}_{\kv_3,\sigma_5}c^{\phantom{*}}_{\kv_3-\qv,\sigma_6} \notag\\
&= - \,
c^{*}_{\kv_2,\sigma_3}c^{\phantom{*}}_{\kv_2+\qv,\sigma_4}
c^{*}_{\kv_1,\sigma_1}c^{*}_{\kv_3,\sigma_5}
c^{\phantom{*}}_{\kv_1,\sigma_2}c^{\phantom{*}}_{\kv_3-\qv,\sigma_6} +
\delta^{\phantom{*}}_{\kv_1,\kv_2}\delta^{\phantom{*}}_{\sigma_2,\sigma_3}
c^{*}_{\kv_1,\sigma_1}c^{\phantom{*}}_{\kv_2+\qv,\sigma_4}
c^{*}_{\kv_3,\sigma_5}c^{\phantom{*}}_{\kv_3-\qv,\sigma_6} \notag\\
&\phantom{=}~ -
\delta^{\phantom{*}}_{\kv_1,\kv_2+\qv}\delta^{\phantom{*}}_{\sigma_1,\sigma_4}
c^{*}_{\kv_2,\sigma_3}c^{\phantom{*}}_{\kv_1,\sigma_2}
c^{*}_{\kv_3,\sigma_5}c^{\phantom{*}}_{\kv_3-\qv,\sigma_6} +
\delta^{\phantom{*}}_{\kv_1,\kv_3}\delta^{\phantom{*}}_{\sigma_2,\sigma_5}
c^{*}_{\kv_2,\sigma_3}c^{\phantom{*}}_{\kv_2+\qv,\sigma_4}
c^{*}_{\kv_1,\sigma_1}c^{\phantom{*}}_{\kv_3-\qv,\sigma_6} \notag\\
&= - \,
c^{*}_{\kv_2,\sigma_3}c^{\phantom{*}}_{\kv_2+\qv,\sigma_4}
c^{*}_{\kv_3,\sigma_5}c^{*}_{\kv_1,\sigma_1}
c^{\phantom{*}}_{\kv_3-\qv,\sigma_6}c^{\phantom{*}}_{\kv_1,\sigma_2} +
\delta^{\phantom{*}}_{\kv_1,\kv_2}\delta^{\phantom{*}}_{\sigma_2,\sigma_3}
c^{*}_{\kv_1,\sigma_1}c^{\phantom{*}}_{\kv_2+\qv,\sigma_4}
c^{*}_{\kv_3,\sigma_5}c^{\phantom{*}}_{\kv_3-\qv,\sigma_6} \notag\\
&\phantom{=}~ -
\delta^{\phantom{*}}_{\kv_1,\kv_2+\qv}\delta^{\phantom{*}}_{\sigma_1,\sigma_4}
c^{*}_{\kv_2,\sigma_3}c^{\phantom{*}}_{\kv_1,\sigma_2}
c^{*}_{\kv_3,\sigma_5}c^{\phantom{*}}_{\kv_3-\qv,\sigma_6} +
\delta^{\phantom{*}}_{\kv_1,\kv_3}\delta^{\phantom{*}}_{\sigma_2,\sigma_5}
c^{*}_{\kv_2,\sigma_3}c^{\phantom{*}}_{\kv_2+\qv,\sigma_4}
c^{*}_{\kv_1,\sigma_1}c^{\phantom{*}}_{\kv_3-\qv,\sigma_6} \notag\\
&= \phantom{-}\,
c^{*}_{\kv_2,\sigma_3}c^{\phantom{*}}_{\kv_2+\qv,\sigma_4}
c^{*}_{\kv_3,\sigma_5}c^{\phantom{*}}_{\kv_3-\qv,\sigma_6}
c^{*}_{\kv_1,\sigma_1}c^{\phantom{*}}_{\kv_1,\sigma_2} \notag\\
&\hspace{0.32cm}+
\delta^{\phantom{*}}_{\kv_1,\kv_2}\delta^{\phantom{*}}_{\sigma_2,\sigma_3}
c^{*}_{\kv_1,\sigma_1}c^{\phantom{*}}_{\kv_2+\qv,\sigma_4}
c^{*}_{\kv_3,\sigma_5}c^{\phantom{*}}_{\kv_3-\qv,\sigma_6} -
\delta^{\phantom{*}}_{\kv_1,\kv_2+\qv}\delta^{\phantom{*}}_{\sigma_1,\sigma_4}
c^{*}_{\kv_2,\sigma_3}c^{\phantom{*}}_{\kv_1,\sigma_2}
c^{*}_{\kv_3,\sigma_5}c^{\phantom{*}}_{\kv_3-\qv,\sigma_6} \notag\\
&\hspace{0.32cm}+
\delta^{\phantom{*}}_{\kv_1,\kv_3}\delta^{\phantom{*}}_{\sigma_2,\sigma_5}
c^{*}_{\kv_2,\sigma_3}c^{\phantom{*}}_{\kv_2+\qv,\sigma_4}
c^{*}_{\kv_1,\sigma_1}c^{\phantom{*}}_{\kv_3-\qv,\sigma_6} -
\delta^{\phantom{*}}_{\kv_1,\kv_3-\qv}\delta^{\phantom{*}}_{\sigma_1,\sigma_6}
c^{*}_{\kv_2,\sigma_3}c^{\phantom{*}}_{\kv_2+\qv,\sigma_4}
c^{*}_{\kv_3,\sigma_5}c^{\phantom{*}}_{\kv_1,\sigma_2}.
\label{eq:gencom}
\end{align}
First, let us focus on the contribution from the hopping amplitude $t$. Therefore, one can take $\overline{\varepsilon}^{\phantom{*}}_{\kv_1,m}c^{*}_{\kv_1,\sigma_1}c^{\phantom{*}}_{\kv_1,\sigma_1}$ instead of $\overline{H}_{m}(\kv_1,\sigma_1)$ in Eq.~\eqref{eq:H3s}, put $\sigma_2=\sigma_1$, $\sigma_4=\sigma_3$, $\sigma_6=\sigma_5$ in the previous calculations and obtain for the first commutator
\begin{align}
\sum_{\substack{\qv \\ m>0 }}
\sum_{\substack{\kv_1, \kv_2, \kv_3 \\ \sigma_1, \sigma_3, \sigma_5}} 
\frac{\overline{\varepsilon}^{\phantom{*}}_{\kv_1,m}
\overline{U}^{\phantom{*}}_{\qv}}{2}
\left[c^{*}_{\kv_1,\sigma_1}c^{\phantom{*}}_{\kv_1,\sigma_1},
c^{*}_{\kv_2,\sigma_3}c^{\phantom{*}}_{\kv_2+\qv,\sigma_3}
c^{*}_{\kv_3,\sigma_5}c^{\phantom{*}}_{\kv_3-\qv,\sigma_5}\right] =
\sum_{\substack{\qv \\ m>0 }}
\sum_{\substack{\kv_2, \kv_3 \\ \sigma_3, \sigma_5}}
\frac{\overline{U}^{\phantom{*}}_{\qv}}{2}
\left( \overline{\varepsilon}^{\phantom{*}}_{\kv_2,m} - \overline{\varepsilon}^{\phantom{*}}_{\kv_2+\qv,m} +
\overline{\varepsilon}^{\phantom{*}}_{\kv_3,m} - \overline{\varepsilon}^{\phantom{*}}_{\kv_3-\qv,m} \right)
c^{*}_{\kv_2,\sigma_3}c^{\phantom{*}}_{\kv_2+\qv,\sigma_3}
c^{*}_{\kv_3,\sigma_5}c^{\phantom{*}}_{\kv_3-\qv,\sigma_5}.
\end{align}
One can see, that the structure of the interaction part did not change after the commutation operation. Indeed, since the Coulomb interaction in our case has the $density\times{}density$ form $\frac12U_{\qv}n_{\qv}n_{\qv}$, where $n_{\qv} = \sum_{\kv\sigma}c^{*}_{\kv,\sigma}c^{\phantom{*}}_{\kv+\qv,\sigma}$. Therefore, the commutation with the hopping term $\overline{\varepsilon}^{\phantom{*}}_{\kv}c^{*}_{\kv,\sigma}c^{\phantom{*}}_{\kv,\sigma}$ will not change the $density\times{}density$ structure of interaction, because it also has the form of the density for $\qv=0$ even if it has the $\kv$--dependent term $\overline{\varepsilon}_{\kv}$ in front of fermionic operators. This important consequence will be very useful for the further calculations.

Then, it is straightforward to see that the contribution to the the second order correction described by Eq.~\eqref{eq:H3s} from hopping amplitude $t$ is equal to
\begin{align}
\tilde{H}_2 = &-\sum_{\substack{\qv \\ m>0 }}
\sum_{\substack{\kv_2, \kv_3 \\ \sigma_3, \sigma_5}}
\frac{\overline{U}_{\qv}}{2m^2}
\left( \overline{\varepsilon}^{\phantom{*}}_{\kv_2,m} - \overline{\varepsilon}^{\phantom{*}}_{\kv_2+\qv,m} +
\overline{\varepsilon}^{\phantom{*}}_{\kv_3,m} - \overline{\varepsilon}^{\phantom{*}}_{\kv_3-\qv,m} \right)
\left( \overline{\varepsilon}^{\phantom{*}}_{\kv_2,-m} - \overline{\varepsilon}^{\phantom{*}}_{\kv_2+\qv,-m} +
\overline{\varepsilon}^{\phantom{*}}_{\kv_3,-m} - \overline{\varepsilon}^{\phantom{*}}_{\kv_3-\qv,-m} \right)
c^{*}_{\kv_2,\sigma_3}c^{\phantom{*}}_{\kv_2+\qv,\sigma_3}
c^{*}_{\kv_3,\sigma_5}c^{\phantom{*}}_{\kv_3-\qv,\sigma_5},
\end{align}
and again has the $density\times{}density$ structure with the two summations $\sum_{\kv_2,\sigma_3}$ and $\sum_{\kv_3,\sigma_5}$.
Here $U_{\qv}$ is the Fourier transform of Coulomb interaction
\begin{align}
U_{\qv} = U_{00} +
2U_{01}\left( \cos{}q_{x} +
2\cos\frac{q_{x}}{2}\cos\frac{\sqrt{3}q_{y}}{2} \right).
\end{align}
Now, let us transform this Hamiltonian back to the real space. For simplicity we use the following notation
\begin{align}
F(\kv_2, \kv_3, \qv) &= \frac{\overline{U}^{\phantom{*}}_{\qv}}{2}\left( \overline{\varepsilon}^{\phantom{*}}_{\kv_2,m} - \overline{\varepsilon}^{\phantom{*}}_{\kv_2+\qv,m} +
\overline{\varepsilon}^{\phantom{*}}_{\kv_3,m} - \overline{\varepsilon}^{\phantom{*}}_{\kv_3-\qv,m} \right)
\left( \overline{\varepsilon}^{\phantom{*}}_{\kv_2,-m} - \overline{\varepsilon}^{\phantom{*}}_{\kv_2+\qv,-m} +
\overline{\varepsilon}^{\phantom{*}}_{\kv_3,-m} - \overline{\varepsilon}^{\phantom{*}}_{\kv_3-\qv,-m} \right) .
\end{align}
Then, transformation looks as ($R=\{r,r',r'',r_1,r_2,r_3,r_4\}$)
\begin{align}
\tilde{H}_{2}(R) &= -\sum_{R}\sum_{m}\sum_{\kv_2, \kv_3, \qv} \frac{F(r, r', r'')}{m^2}
c^{*}_{r_1,\sigma_2}c^{\phantom{*}}_{r_2,\sigma_2}
c^{*}_{r_3,\sigma_3}c^{\phantom{*}}_{r_4,\sigma_3}
e^{i\kv_2r+i\kv_3r'+i\qv{}r''}e^{i\kv_2r_1-i(\kv_2+\qv)r_2+i\kv_3r_3-i(\kv_3-\qv)r_4}\notag\\
&=-\sum_{R}\sum_{m}\sum_{\kv_2, \kv_3, \qv} \frac{F(r, r', r'')}{m^2}
c^{*}_{r_1,\sigma_2}c^{\phantom{*}}_{r_2,\sigma_2}
c^{*}_{r_3,\sigma_3}c^{\phantom{*}}_{r_4,\sigma_3}
e^{i\kv_2(r+r_1-r_2)}e^{i\kv_3(r'+r_3-r_4)}e^{i\qv(r''-r_2+r_4)}\notag\\
&=-\sum_{R}\sum_{m} \frac{F(r, r', r_1+r-r_3-r')}{m^2}
c^{*}_{r_1,\sigma_2}c^{\phantom{*}}_{r_1+r,\sigma_2}
c^{*}_{r_3,\sigma_3}c^{\phantom{*}}_{r_3+r',\sigma_2},
\end{align}
where
\begin{align}
F(r,r',r'') =  \sum_{\kv_2, \kv_3, \qv} F(\kv_2, \kv_3, \qv)
e^{-i\kv_2r-i\kv_3r'-i\qv{}r''}.
\end{align}
Since, in general, the Coulomb potential decays as $1/r$ and the Fourier transform of the function $F_{U}(\kv_2, \kv_3, \qv)$ also rapidly decays with the distance, we consistently restrict ourselves to the one-site and nearest-neighbour two-site approximation in the same way as it was done in~\cite{Itin2}. These terms give the main contribution to the interaction, so these approximation is sufficient. Therefore, there are four possibilities to obtain such terms:\\
{\bf a)} $r_1=r_3+r'=r_1+r=r_3=i$, so $r=r'=r''=0$,\\
{\bf b)} $r_1=r_1+r=i$ and $r_3=r_3+r'=j$, so $r=r'=0$ and $r''=r_1-r_3=a_0$,\\
{\bf c)} $r_1=r_3=i$ and $r_1+r=r_3+r'=j$, so $r=r'=a_0$ and $r''=0$,\\
{\bf d)} $r_1=r_3+r'=i$ and $r_1+r=r_3=j$, so $r=-r'=a_0$ and $r''=r=a_0$.\\
Then, the second-order term $\tilde{H}_{2}$ for the all four cases is given by (for all possible values of $m$)
\begin{align}
&\tilde{H}^{\bf a}_2 = -\sum_{i}\sum_{m>0}
\sum_{\substack{\kv_2, \kv_3, \qv\\ \sigma \sigma'}}
\frac{F(\kv_2, \kv_3, \qv)}{m^2}\,
c^{*}_{i,\sigma}c^{\phantom{*}}_{i,\sigma}
c^{*}_{i,\sigma'}c^{\phantom{*}}_{i,\sigma'}
= -\sum_{i}\sum_{\sigma \sigma'}\sum_{m>0}
8\,\overline{t}^2{\cal J}^2_{m}(Z) \frac{(\overline{U}_{00}-\overline{U}_{\av{ij}})}{m^2}\,
n^{\phantom{*}}_{i,\sigma}
n^{\phantom{*}}_{i,\sigma'},\label{Coulomb00}\\
&\tilde{H}^{\bf b}_2 = -\sum_{\av{ij}}\sum_{m>0}
\sum_{\substack{\kv_2, \kv_3, \qv\\ \sigma \sigma'}}
\frac{F(\kv_2, \kv_3, \qv)}{m^2}\,
c^{*}_{i,\sigma}c^{\phantom{*}}_{i,\sigma}
c^{*}_{j,\sigma'}c^{\phantom{*}}_{j,\sigma'}\,
e^{-i\qv{}a}
= -\sum_{\av{ij}}\sum_{\sigma \sigma'}\sum_{m>0}
2\,\overline{t}^2{\cal J}^2_{m}(Z) \frac{(4\overline{U}_{\av{ij}}-\overline{U}_{00})}{m^2}\,
n^{\phantom{*}}_{i,\sigma}
n^{\phantom{*}}_{j,\sigma'},\label{Coulomb01}\\
&\tilde{H}^{\bf c}_2 = -\sum_{\av{ij}}\sum_{m>0}
\sum_{\substack{\kv_2, \kv_3, \qv\\ \sigma \sigma'}}
\frac{F(\kv_2, \kv_3, \qv)}{m^2}\,
c^{*}_{i,\sigma}c^{\phantom{*}}_{j,\sigma}
c^{*}_{i,\sigma'}c^{\phantom{*}}_{j,\sigma'}\,
e^{-i\kv_2a-i\kv_3a} = -
\sum_{\av{ij}}\sum_{\sigma \sigma'}\sum_{m>0}
2\,\overline{t}^2(-1)^{m}{\cal J}^2_{m}(Z) \frac{(\overline{U}_{00}-\overline{U}_{\av{ij}})}{m^2}\,
d^{*}_{i}d^{\phantom{*}}_{j},
\label{doublons}\\
&\tilde{H}^{\bf d}_2 = -\sum_{\av{ij}}\sum_{m>0}
\sum_{\substack{\kv_2, \kv_3, \qv\\ \sigma \sigma'}}
\frac{F(\kv_2, \kv_3, \qv)}{m^2}\,
c^{*}_{i,\sigma}c^{\phantom{*}}_{j,\sigma}
c^{*}_{j,\sigma'}c^{\phantom{*}}_{i,\sigma'}\,
e^{-i\kv_2a+i\kv_3a-i\qv{}a} =
-\sum_{\av{ij}}\sum_{\sigma \sigma'}\sum_{m>0}
2\,\overline{t}^2{\cal J}^2_{m}(Z) \frac{(\overline{U}_{00}-\overline{U}_{\av{ij}})}{m^2}\,
c^{*}_{i,\sigma}c^{\phantom{*}}_{i,\sigma'}
c^{*}_{j,\sigma'}c^{\phantom{*}}_{j,\sigma}\label{eq:Jd},
\end{align}
where we considered the case of circularly polarized laser field ($Z=eA_{x}a_0=eA_{y}a_0$, $\phi=0$) for simplicity. Here, the first and the second terms $\tilde{H}^{\bf a}_{2}$ and $\tilde{H}^{\bf b}_{2}$ give a correction to the local and nearest-neighbour Coulomb interaction respectively. These corrections come with the factor $\lambda^2$ and therefore are negligibly small with respect to the Coulomb interactions that are already presented in the main term $\tilde{H}_{0}$ of the effective Hamiltonian. Contrary to them, the fourth term $\tilde{H}^{\bf d}_{2}$ induced by the laser field gives a very important contribution $J^{\rm D}_{\rm ind}$ to the direct exchange that comes from the Coulomb interaction as we stressed in the beginning of this Section, and therefore, is much larger then a correction that might appear directly from $J^{\rm D}$. It is worth mentioning, that the term described by Eq.~\eqref{doublons} is new and this type of interaction does not exist in the initial Hamiltonian. This term describes kinetics of doublons, where $d^{*}_{i} = c^{*}_{i,\sigma}c^{*}_{i,-\sigma}$ and $d^{\phantom{*}}_{i} = c_{i,-\sigma}c_{i,\sigma}$ are the creation and annihilation operators of doublons. As it is shown in the Ref.~\onlinecite{Itin2}, this term does not contribute to the exchange interaction, and we will not consider it here.

It is not surprising, that exactly the $\overline{\varepsilon}^{\phantom{*}}_{\kv_1,m}c^{*}_{\kv_1,\sigma_1}c^{\phantom{*}}_{\kv_1,\sigma_1}$ part of $\overline{H}_{m}(\kv_1,\sigma_1)$ in Eq.~\eqref{eq:H3s} generates an effective exchange interaction $J^{\rm D}_{\rm ind}$. As it was discussed above, the commutation with the hopping term that has the density structure does not change the $density\times{}density$ structure of interaction, and the final result for the Eq.~\eqref{eq:H3s} in this case has the $density\times{}density$ structure as well. The exchange interaction enters the Heisenberg Hamiltonian as $J_{ij}\hat{\bf S}_{i}\hat{\bf S}_{j}$. The $\hat{\bf S}^2$ operator, that also can be written as $\hat{S}^{x}\hat{S}^{x}+\hat{S}^{y}\hat{S}^{y}+\hat{S}^{z}\hat{S}^{z}$ is proportional to the $density\times{}density$ term $nn$, therefore the obtained result was expectable.\\

Now let us study whether it is possible to induce the anisotropic DMI interaction by the laser field similarly to the case of the exchange interaction. DMI enters the Heisenberg Hamiltonian as ${\bf D}_{ij}\,[\hat{\bf S}_{i}\times{}\hat{\bf S}_{j}]$ and can be rewritten as $D^{x}_{ij}\,(\hat{S}^{y}_{i}\hat{S}^{z}_{j}-~\hat{S}^{z}_{i}\hat{S}^{y}_{j}) + D^{y}_{ij}\,(\hat{S}^{z}_{i}\hat{S}^{x}_{j}-~\hat{S}^{x}_{i}\hat{S}^{z}_{j})$, since the ${\bf D}_{ij}$ is determined by the ${\bf \Delta}_{ij}$ that has only $x$ and $y$ components. Therefore, the DMI has the structure of $\hat{S}^{z}$ operator that couples to the $x$ or $y$ component of the spin operator $\hat{\bf S}$. One can see that similarly to the hopping term, the spin-orbit term $i\sum_{\kv,\,\sigma\sigma'} \boldsymbol{f}^{\phantom{*}}_{\kv,m} c^{*}_{\kv\sigma}
\boldsymbol{\sigma}^{\phantom{*}}_{\sigma\sigma'}
c^{\phantom{*}}_{\kv\sigma'}$ in the initial Hamiltonian~\eqref{eq:Hamfr} has the form of spin operators $\hat{S}^{x(y)}_{\qv=0}$ ($\hat{S}^{x(y)}_{\qv} =~ \frac12\sum_{\kv,\sigma\sigma'}c^{*}_{\kv,\sigma}\sigma^{x(y)}_{\sigma\sigma'}c^{\phantom{*}}_{\kv+\qv,\sigma'}$), even if there is again a $\kv$--dependent coefficient $\boldsymbol{f}_{\kv}$ in front of fermionic operators. Since the kinetic part of DMI is determined as ${\bf D} = \frac{4t\mathbf{\Delta}}{U}$ and according to the structure of DMI discussed above, one could expect, that the only one possibility to induce a direct DMI is described by the following contribution to the second order correction $\tilde{H}_2$
\begin{align}
\tilde{H}_2 = \sum_{\substack{\qv \\ m\neq0}}
\sum_{\substack{\kv_1, \kv_2, \kv_3, \kv_4 \\ \sigma_1, \sigma_3, \sigma_5, \sigma, \sigma'}}
\frac{\left[\left[\overline{\varepsilon}^{\phantom{*}}_{\kv_1,m}c^{*}_{\kv_1,\sigma_1}c^{\phantom{*}}_{\kv_1,\sigma_1},
\frac12\overline{U}^{\phantom{*}}_{\qv}c^{*}_{\kv_2,\sigma_3}c^{\phantom{*}}_{\kv_2+\qv,\sigma_3}
c^{*}_{\kv_3,\sigma_5}c^{\phantom{*}}_{\kv_3-\qv,\sigma_5}
\right],
i\overline{\boldsymbol{f}}^{\phantom{*}}_{\kv_4,-m}c^{*}_{\kv_4\sigma}
\boldsymbol{\sigma}^{\phantom{*}}_{\sigma\sigma'}
c^{\phantom{*}}_{\kv_4\sigma'}\right]}{m^2}.
\end{align}
Let us study the case of $x$ component $i\sum_{\kv,\sigma} \overline{f}^{x}_{\kv,m} c^{*}_{\kv,\sigma}c^{\phantom{*}}_{\kv,-\sigma}$ of the spin-orbit term (for the $y$ component the calculations are similar). Using the Eq.~\eqref{eq:gencom} and the fact that the commutation of the interaction with the hopping term does not change the interaction, one can get the following result \begin{align}
\tilde{H}_2 = -\sum_{\substack{\qv \\ m\neq0}}
\sum_{\substack{\kv_2, \kv_3 \\ \sigma_3, \sigma_5,}}
\frac{\overline{U}^{\phantom{*}}_{\qv}}{2m^2}
\left( \overline{\varepsilon}^{\phantom{*}}_{\kv_2,m} - \overline{\varepsilon}^{\phantom{*}}_{\kv_2+\qv,m} +
\overline{\varepsilon}^{\phantom{*}}_{\kv_3,m} - \overline{\varepsilon}^{\phantom{*}}_{\kv_3-\qv,m} \right)
&\left(\overline{f}^{x}_{\kv_2,-m}
c^{*}_{\kv_2,-\sigma_3}c^{\phantom{*}}_{\kv_2+\qv,\sigma_3}
c^{*}_{\kv_3,\sigma_5}c^{\phantom{*}}_{\kv_3-\qv,\sigma_5}
-\overline{f}^{x}_{\kv_2+\qv,-m}
c^{*}_{\kv_2,\sigma_3}c^{\phantom{*}}_{\kv_2+\qv,-\sigma_3}
c^{*}_{\kv_3,\sigma_5}c^{\phantom{*}}_{\kv_3-\qv,\sigma_5}\right. \notag\\
&\hspace{-0.15cm}\left.+\overline{f}^{x}_{\kv_3,-m}
c^{*}_{\kv_2,\sigma_3}c^{\phantom{*}}_{\kv_2+\qv,\sigma_3}
c^{*}_{\kv_3,-\sigma_5}c^{\phantom{*}}_{\kv_3-\qv,\sigma_5}
-\overline{f}^{x}_{\kv_3-\qv,-m}
c^{*}_{\kv_2,\sigma_3}c^{\phantom{*}}_{\kv_2+\qv,\sigma_3}
c^{*}_{\kv_3,\sigma_5}c^{\phantom{*}}_{\kv_3-\qv,-\sigma_5}
\right).
\end{align}
Unfortunately, the obtained result has the form of $x$ component $\hat{S}^{x}$ of the spin operator that is coupled to the density $n$, but not to the $\hat{S}^{z}$ operator. Therefore, this term does not contribute to the DMI interaction and does not affect the exchange interaction. This result is consistent with the logic presented above. Indeed, the commutation of the density-like term $\varepsilon_{\kv}$ with interaction and with spin-like term $\boldsymbol{f}_{\kv}$ will produce only $density\times{}\hat{S}^{x(y)}$ contribution, but not an $\hat{S}^{z}\times{}\hat{S}^{x(y)}$-like terms. Therefore, an anisotropic DMI interaction can not be induced by the laser field.  \\

Finally, the effective Hamiltonian can be written as
\begin{align}
{\cal H} = \overline{\cal H}\delta{}E \simeq \left(\tilde{H}_0 + \lambda^2\tilde{H}_2\right)\delta{}E = &\sum_{\av{ij},\,\sigma\sigma'}c^{*}_{i\sigma}\left(t'
\,\delta^{\phantom{*}}_{\sigma\sigma'} + i
\boldsymbol{\Delta}'_{ij}\,
\boldsymbol{\sigma}^{\phantom{*}}_{\sigma\sigma'}
\right)c^{\phantom{*}}_{j\sigma'}
+\sum_{i}U^{\phantom{*}}_{00}\,
n^{\phantom{*}}_{i\uparrow}n^{\phantom{*}}_{i\downarrow} + \\ &\frac12\sum_{\av{ij},\,\sigma\sigma'}U^{\phantom{*}}_{\av{ij}}\,
n^{\phantom{*}}_{i\sigma}n^{\phantom{*}}_{j\sigma'}
-\frac12\sum_{\av{ij},\,\sigma\sigma'}
J'^{\rm D}_{\av{ij}}\,
c^{*}_{i\sigma}c^{\phantom{*}}_{i,\sigma'}
c^{*}_{j,\sigma'}c^{\phantom{*}}_{j\sigma}, \notag
\end{align}
where direct exchange interaction is also renormalized by the laser field and is equal to $J'^{\rm D}_{\av{ij}} = J^{\rm D}_{\av{ij}} + J^{\rm D}_{\rm ind}$, where
\begin{align}
J^{\rm D}_{\rm ind} = \frac{4\,t^2}{\Omega^2}\sum_{m>0}{\cal J}^2_{m}(Z) \frac{(U_{00}-U_{\av{ij}})}{m^2},
\end{align}
The renormalized hopping amplitude and the spin-orbit coupling vector were determined above.

\subsection{Effective time-independent Hamiltonian in the high-frequency representation for $t_1-t_{2,3}$ model}
Performing the similar transformations, one can get the Fourier transform of the kinetic part of the initial time-dependent Hamiltonian as $\varepsilon^{\phantom{*}}_{\kv,m} = \varepsilon^{1}_{\kv,m} + \varepsilon^{2(3)}_{\kv,m}$, where
\begin{align}
\varepsilon^{1}_{\kv,m} &=
2t_1{\cal J}_{m}(Z)\left[\cos(k_{x}-m\pi/2)
+\cos(k_{x}/2+k_{y}\sqrt{3}/2)\,
e^{im5\pi/6}+\cos(k_{x}/2-k_{y}\sqrt{3}/2)\,
e^{im\pi/6}\right], \\
\varepsilon^{2}_{\kv,m} &=
2t_2{\cal J}_{m}(\sqrt{3}Z)\left[\cos(k_{y}\sqrt{3}+m\pi/2)\,e^{-im\pi/2}
+\cos(k_{x}3/2+k_{y}\sqrt{3}/2)\,
e^{im2\pi/3}+\cos(k_{x}3/2-k_{y}\sqrt{3}/2)\,
e^{im\pi/3}\right],\\
\varepsilon^{3}_{\kv,m} &=
2t_3{\cal J}_{m}(2Z)\left[\cos(2k_{x}-m\pi/2)
+\cos(k_{x}+k_{y}\sqrt{3})\,
e^{im5\pi/6}+\cos(k_{x}-k_{y}\sqrt{3})\,
e^{im\pi/6}\right].
\end{align}
Here $t_1$ is the nearest-neighbor hopping amplitude, $t_{2,3}$ are the nearest-NN hoppings, and we again consider the case of the circularly polarized light ($Z=eA_{x}a_0=eA_{y}a_0$, $\phi=0$). \\
Similarly to the case presented above, there is only one correction~\eqref{eq:Jd} that matters for the magnetic properties of the considered model. Then the correction $\tilde{H}_{3}$ for the effective Hamiltonian reads
\begin{align}
&\tilde{H}^{\bf d}_2 = -\sum_{\av{ij}}\sum_{m>0}
\sum_{\substack{\kv_2, \kv_3, \qv\\ \sigma_2, \sigma_3}}
\frac{F(\kv_2, \kv_3, \qv)}{m^2}\,
c^{*}_{i,\sigma_2}c^{\phantom{*}}_{j,\sigma_2}
c^{*}_{j,\sigma_3}c^{\phantom{*}}_{i,\sigma_3}\,
e^{-i\kv_2r_{i}+i\kv_3r_{i}-i\qv{}r_{i}}
\label{eq:Jd123}.
\end{align}
Since in the $t_1-t_{2,3}$ model we additionally included the next-NN hopping processes, we will also consider the next-NN two-site contributions $r_{2}=\sqrt{3}a_0$, $r_{3}=2a_0$ here in addition to the nearest-neighbor case of $r_1=a_0$, and we again define
\begin{align}
F(\kv_2, \kv_3, \qv) &= \frac{\overline{U}^{\phantom{*}}_{\qv}}{2} \left( \overline{\varepsilon}^{\phantom{*}}_{\kv_2,m} - \overline{\varepsilon}^{\phantom{*}}_{\kv_2+\qv,m} +
\overline{\varepsilon}^{\phantom{*}}_{\kv_3,m} - \overline{\varepsilon}^{\phantom{*}}_{\kv_3-\qv,m} \right)
\left( \overline{\varepsilon}^{\phantom{*}}_{\kv_2,-m} - \overline{\varepsilon}^{\phantom{*}}_{\kv_2+\qv,-m} +
\overline{\varepsilon}^{\phantom{*}}_{\kv_3,-m} - \overline{\varepsilon}^{\phantom{*}}_{\kv_3-\qv,-m} \right)
\end{align}
and the Fourier transform of Coulomb interaction as
\begin{align}
U_{\qv} = U_{00} +
2U_{01}\left( \cos{}q_{x} +
2\cos\frac{q_{x}}{2}\cos\frac{\sqrt{3}q_{y}}{2} \right) +
2U_{02}\left( \cos\sqrt{3}q_{y} +
2\cos\frac{3q_{x}}{2}\cos\frac{\sqrt{3}q_{y}}{2} \right) +
2U_{03}\left( \cos2q_{x} +
2\cos{}q_{x}\cos\sqrt{3}q_{y} \right).
\end{align}
\\
The final effective Hamiltonian now reads
\begin{align}
&{\cal H} = \sum_{\av{ij},\,\sigma}t'_{1}\, c^{*}_{i\sigma}\,c^{\phantom{*}}_{j\sigma} +
\sum_{\av{\av{ij}},\,\sigma}t'_{2,3}\,
c^{*}_{i\sigma}\,c^{\phantom{*}}_{j\sigma} +
\sum_{i}U^{\phantom{*}}_{00}\, n^{\phantom{*}}_{i\uparrow}n^{\phantom{*}}_{i\downarrow}
+\frac12\sum_{ij,\,\sigma\sigma'}\left(U^{\phantom{*}}_{\av{ij}}+U^{\phantom{*}}_{\av{\av{ij}}}\right)
n^{\phantom{*}}_{i\sigma}n^{\phantom{*}}_{j\sigma'}
-\frac12\sum_{ij,\,\sigma\sigma'}
J'^{\rm D}_{ij}\,
c^{*}_{i\sigma}c^{\phantom{*}}_{i,\sigma'}
c^{*}_{j,\sigma'}c^{\phantom{*}}_{j\sigma},
\end{align}
where the renormalized hopping amplitudes are $t'_{1} = t_1\,{\cal J}_0(Z)$, $t'_{2} = t_2\,{\cal J}_0(\sqrt{3}Z)$, $t'_{3} = t_3\,{\cal J}_0(2Z)$, and the renormalized direct exchange interaction can be obtained from the Eq.~\eqref{eq:Jd123} as
\begin{align}
J'^{\rm D}_{01} &= J^{\rm D}_{01} + \frac{4\,t_1^2}{\Omega^2}\sum_{m>0}{\cal J}^2_{m}(Z) \frac{(U_{00}-U_{01})}{m^2},\\
J'^{\rm D}_{02} &= \frac{4\,t_2^2}{\Omega^2}\sum_{m>0}{\cal J}^2_{m}(\sqrt{3}Z) \frac{(U_{00}-U_{02})}{m^2},\\
J'^{\rm D}_{03} &= \frac{4\,t_3^2}{\Omega^2}\sum_{m>0}{\cal J}^2_{m}(2Z) \frac{(U_{00}-U_{03})}{m^2}.
\end{align}

\end{document}